\documentclass[sigconf,nonacm]{acmart}

\usepackage[subtle]{savetrees}
\usepackage{comment}
\usepackage{xcolor}
\usepackage{xspace}
\usepackage{soul} 
\usepackage{subcaption} 
\usepackage[font=small,skip=0pt]{caption}

\newcommand{\karthika}[1]{\textcolor{purple}{[(Karthika): #1]}}
\newcommand{\roberto}[1]{\textcolor{red}{[(Roberto): #1]}}

\newcommand{\sysname}{{{\fontfamily{lmss}\selectfont IXmon}}\xspace}

\AtBeginDocument{%
	\providecommand\BibTeX{{%
			\normalfont B\kern-0.5em{\scshape i\kern-0.25em b}\kern-0.8em\TeX}}}

\begin{document}
	
	\title{IXmon: Detecting and Analyzing DRDoS Attacks at Internet Exchange Points}
	
	\author{Karthika Subramani}
	\affiliation{%
		\institution{University of Georgia}
		}
	\email{ks54471@uga.edu}
	
	\author{Roberto Perdisci}
	\affiliation{%
		\institution{University of Georgia and Georgia Institute of Technology }
		}
	\email{perdisci@cs.uga.edu}
	
	\author{Maria Konte}
	\affiliation{%
		\institution{Georgia Institute of Technology}
		}
	\email{mkonte@gatech.edu}
	
	
    \begin{abstract}
    Distributed reflective denial of service (DRDoS) attacks are a popular
    choice among adversaries. In fact, one of the largest DDoS attacks ever
    recorded, reaching a peak of 1.3Tbps against GitHub, was a memcached-based
    DRDoS attack. More recently, a record-breaking 2.3Tbps attack against Amazon
    AWS was due to a CLDAP-based DRDoS attack. Although reflective attacks have
    been known for years, DRDoS attacks are unfortunately still popular and
    largely unmitigated.

    In this paper, we study in-the-wild DRDoS attacks observed from a large
    Internet exchange point (IXP) and provide a number of security-relevant
    measurements and insights. To enable this study, we first developed
    \sysname, an open-source DRDoS detection system specifically designed for
    deployment at large IXP-like network connectivity providers and peering
    hubs. We deployed \sysname at Southern Crossroads (SoX), an IXP-like hub
    that provides both peering and upstream Internet connectivity services to
    more than 20 research and education (R\&E) networks in the South-East United
    States. In a period of about 21 months, \sysname detected more than 900
    DRDoS attacks towards 31 different victim ASes. An analysis of the
    real-world DRDoS attacks detected by our system shows that most DRDoS
    attacks are short lived, lasting only a few minutes, but that large-volume,
    long-lasting, and highly-distributed attacks against R\&E networks are not
    uncommon. We then use the results of our analysis to discuss possible attack
    mitigation approaches that can be deployed at the IXP level, before the
    attack traffic overwhelms the victim's network bandwidth.
\end{abstract}

	
	\maketitle

\section{Introduction}

Large-scale distributed denial of service (DDoS) attacks pose an imminent threat
to the availability of critical Internet-based operations~\cite{Kang2013}, and
have become part of sophisticated cyber-warfare arsenals~\cite{EstoniaCyberWar}.
DDoS attacks can take many different forms~\cite{Mirkovic2004}, and leverage
weaknesses that span from the application-layer to the physical-layer. In
particular, recent incidents have demonstrated that {\em bandwidth exhaustion}
DDoS attacks are capable of bringing down even the most well-provisioned
Internet services, such as highly popular websites (e.g., Twitter, Netflix,
etc.) and cybersecurity services~\cite{DynDoS, SpamhausDoS, KrebsDoS,
DigitalAttackMap}. 
Among bandwidth exhaustion attacks, distributed
reflective denial of service (DRDoS) attacks are a popular choice among
adversaries. In fact, one of the largest DDoS attacks ever recorded, reaching a
peak of 1.3Tbps against GitHub, was a memcached-based DRDoS
attack~\cite{cloudflare-famousDDoS}. More recently, a record-breaking 2.3Tbps
attack against Amazon AWS was due to a CLDAP-based DRDoS attack~\cite{aws-DRDoS}.

Although
reflective attacks have been known for years~\cite{AmplificationHell} and could be
mitigated in part by filtering/throttling traffic to/from some UDP services
(e.g., filtering memcached traffic at the edge of a
network~\cite{Cloudflare-memcached}), DRDoS attacks are unfortunately still
popular~\cite{NSFocus} and largely unmitigated. At the same time, while some
information about DRDoS attacks can be found in blog posts or white papers from
security vendors (e.g., \cite{Cloudflare-reflections}), there is a lack of systematic
studies that provide an in-depth measurement of the properties of {\em in-the-wild} DRDoS attacks, such as
occurrence frequency, the distribution of their sources, duration, volume, targets, and what
mitigation steps could be applied to combat them.

In this paper, we aim to partly fill this gap by studying real-world DRDoS attacks
observed from a large Internet exchange point (IXP)\footnote{Whereas others may
define IXPs purely as facilitating public peering among networks, we refer to
IXPs in a broader sense as hubs that facilitate both peering and commercial
connectivity (e.g., transit) services.}. IXPs are high-density
peering and connectivity hubs that provide infrastructure used by autonomous systems (ASes) to
interconnect with each other (e.g., public or private peering and other
connectivity agreements). Because IXPs provide an increasingly large portion
of the global Internet infrastructure used by ASes to exchange traffic, they can
play a key role in detecting and mitigating DDoS attacks.

To enable this study, we first develop \sysname, an open-source DRDoS detection system
specifically designed for deployment at large IXP-like network connectivity
providers and peering hubs. While there exist several DDoS detection and mitigation 
solutions, such as {\em traffic scrubbing} services~\cite{Cloudflare,
Akamai}, these are typically expensive third-party commercial services. In
addition, they are not designed for detecting DRDoS attacks at IXPs, and are
instead more focused on inline DDoS traffic detection and traffic filtering. On the
other hand, our \sysname system is fully open-source, can be deployed at large IXPs,
and can also be used to enable IXP-based DDoS mitigations. In fact, \sysname's goal is not only to detect the
occurrence of a DRDoS attack very early after its inception, but also to
identify ASes that host the reflectors used in the attack. This capability could be used to enable the filtering of DRDoS
attack traffic at the IXP level before it is
routed to the victim, thus preventing the victim's network bandwidth from being
exhausted.

\begin{figure*}[t]
    \centering
    \includegraphics[scale=0.4]{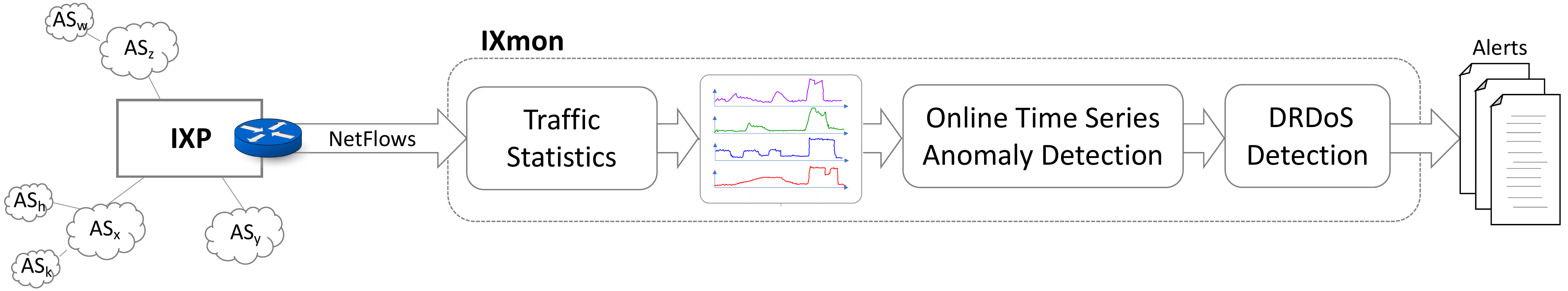}
    \caption{\sysname system overview}
    \label{fig:ixp-overview}
    \vspace{-10pt}
\end{figure*}

\begin{figure}[t]
    \centering
    \includegraphics[width=\linewidth]{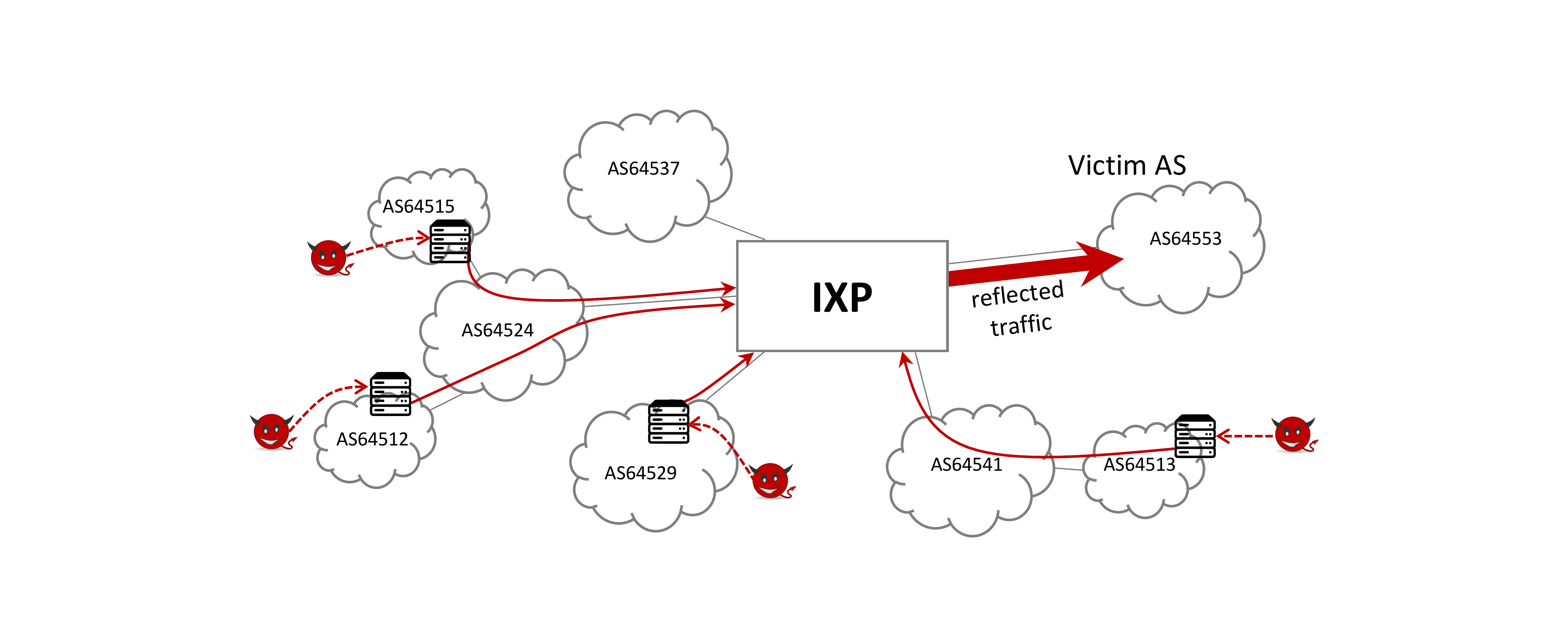}
    \caption{Example of reflection attack traffic flowing through an IXP}
    \label{fig:ixp-reflection}
\end{figure}

Figure~\ref{fig:ixp-overview} provides an overview of our \sysname system,
whereas Figure~\ref{fig:ixp-reflection} shows an example of how reflected
traffic belonging to a DRDoS attack
may traverse an IXP's fabric to reach the victim network.
To detect DRDoS attacks, \sysname takes in input network flow summaries (e.g., using Cisco's NetFlow v9
format~\cite{NetFlowv9}), which report flow statistics for all traffic
from any source IP to a any destination IP that crosses the IXP. Because \sysname aims to detect DRDoS attacks, we focus
on UDP flows whose source port is associated with services that can be
abused for amplification attacks, such as {\tt DNS}, {\tt NTP}, {\tt memcached},
{\tt CLDAP}, etc.~\cite{AmplificationHell} (see Section~\ref{sec:ixmon} for a complete
list). Given a specific service (e.g., {\tt
memcached}), we then aggregate all related UDP flows directed to each destination AS and compute the overall traffic
volume of all flows belonging to the same $(service, dstAS)$ pair. We update
these aggregate flow statistics in an online fashion at regular (small) time intervals,
and perform online time series anomaly detection to detect highly anomalous
increases in traffic volume. Finally, every time an anomalous traffic volume
increase is detected for a $(service, dstAS)$ pair, we pass this information to
the DRDoS detection module, which applies additional checks to filter out
possible false positives and only issue an alert for events that are highly
likely associated with actual DRDoS attacks. Additionally, the DRDoS detection
module identifies the source ASes involved in an attack, and ranks them
according to the attack traffic volume they contribute. By knowing the UDP
source port number, the destination AS (i.e., the victim network), and the
source ASes that contribute the highest amount of attack traffic, an IXP could
then deploy traffic filtering rules to mitigate the attack in its very early
stages. In fact, this filtering rule deployment process could be automated by
automatically deriving BGPFlowSpec rules~\cite{bgp_flowspec} from IXmon's alerts.

We have deployed \sysname at Southern Crossroads (SoX)~\cite{SoX}, an IXP-like hub that provides both
peering and upstream Internet connectivity services to more than 20 research and
education (R\&E) networks in the South-East United States. In a period of about
21 months, \sysname detected more than 900 DRDoS attacks towards 31 different victim
ASes. In Section~\ref{sec:measurements}, we study the characteristics of these attacks and present a number of
insights regarding their duration and intensity, what services are
most abused, what networks are more often targeted, and whether the victim
networks took action to mitigate the attacks.

In summary, we make the following contributions:
\begin{itemize}
    \item We present \sysname, an open-source DRDoS detection system
    specifically designed to be deployed at large IXP-like peering and
    connectivity hubs. IXmon uses online time series anomaly detection and
    traffic analysis methods to detect both the victims and sources of
    real-world DRDoS attacks.
    \item We deploy \sysname at a large IXP-like R\&E peering and connectivity
    hub located in the South-East United Sates for a period of about 21 months,
    where we detected a large number and variety of real-world DRDoS attacks in
    near real time.
    \item We analyze the real-world DRDoS attacks detected by our system and
    report a number of security-relevant measurements and insights. For
    instance, we show that most DRDoS attacks are short lived, lasting only a
    few minutes, but that large-volume, long-lasting, and highly-distributed
    attacks against R\&E networks are not uncommon.
\end{itemize}

\section{Background on IXPs}

IXPs have been traditionally established as infrastructures
that primarily offer peering services. The primary role of an
IXP is to serve as a physical exchange point to facilitate the exchange of
Internet traffic between different autonomous systems (ASes). The minimum number
of ASes that interconnect at an IXP should be at least three and there must be a
clear and open policy for other ASes to join~\cite{description}. The ASes
interconnect through a shared switching fabric that the IXPs offer. This
interconnection infrastructure can vary widely in complexity. Some
infrastructures can be very simple and minimal (as a single switch), or very
complex (as a large scale distributed infrastructure that includes remote
peering)~\cite{DrPeering}.

Since their initial establishment, the role of
IXPs has been evolving along with their offered services. Some services are
offered as free value-added services and others are paid services. Many IXPs
offer both
public peering and private peering, multi-lateral and bi-lateral peering, data
center services, multiple network management and other services including route
servers, SDN-based network management, traffic engineering, and traffic
blackholing. 

IXPs have been recently further evolving towards becoming major
peering and connectivity hubs, claiming a central role as part of the Internet's
core infrastructures \cite{chatzis2013there,
richter2014peering, ager2012anatomy}. There are currently more than 500 IXPs
worldwide. Both their membership growth and the traffic growth show their
dynamic and evolving role in the Internet ecosystem. Some of the largest IXPs
have more than 700 members, while they carry as much traffic as some of the
largest global Tier-1 ISPs.
In this work, we refer to IXPs in this latter broader sense, as exchange points
in which multiple ASes peer with each other and can connect to upstream
Internet connectivity services.


\section{IXmon System}
\label{sec:ixmon}

In this section, we describe how \sysname's components work, following the high-level overview
shown in Figure~\ref{fig:ixp-overview}.

\vspace{3pt}
\noindent
{\bf Approach Overview}:
Given the traffic towards a specific AS, $A$, to detect DRDoS attacks against
$A$ we look for the following factors:
\begin{enumerate}
    \item Focus on traffic coming from a UDP source port typically associated
    with a service that can be abused for attack amplification.
    \item For each of those source ports, has the traffic volume towards $A$
    increased in a highly anomalous way?
    \item Is the anomalous traffic distributed
    across several contributing source ASes?
\end{enumerate}

As an example, assume that a destination AS $A$ usually receives very low
amounts of traffic from source port UDP 123, which is typically associated with
the NTP service. We monitor all traffic from port UDP 123 that flows towards $A$
through the IXP's fabric. All of a sudden, at time $t$ we detect a spike in
incoming NTP traffic, and notice that several different source AS numbers are
contributing in a coordinated way to this traffic spike. This scenario meets the
``recipe'' for a DRDoS attack, which \sysname aims to detect automatically. In
the following, we explain how we translate the above high-level
approach into a concrete DRDoS detection system.

\subsection{Aggregate Traffic Statistics}
\label{sec:agg_traffic}

\sysname is designed to monitor network traffic at large real-world IXP-like
peering and connectivity hubs. Due to the sheer amount of traffic observed from
such a vantage point, efficiency is a high priority goal. In particular, memory
consumption is a main concern, given the large amount of network traffic
statistics that we need to track over all possible targets and sources of DRDoS
attacks visible from an IXP. To this end, our first step is to condense detailed
information about network flows crossing the IXP into {\em traffic sketches}
containing aggregate traffic statistics.

\sysname receives network flow statistics as input. While our
current implementation supports and has been tested only on Cisco NetFlow
versions 9 and 10~\cite{NetFlowL2}, it is designed to also support other formats,
including sFlow~\cite{sFlow}. For simplicity, in the following we will simply use
the term {\em flow} to refer to a network flow in NetFlow format. While
NetFlow flows include many details about how the related network packets
traversed the IXP (e.g., including the network interfaces involved in routing
the flow), we will only refer to the properties that are used by our system.

Let the tuple 
\begin{equation*}
f_i=(srcIP_i, srcPort_i, dstIP_i, dstPort_i, protocol_i, packets_i, bytes_i)
\end{equation*}
represent a network flow, where $packets_i$ and $bytes_i$ represent the number of
packets and overall number of bytes sent from the source to the destination
IPs/ports that have been ``captured'' by flow $f_i$. The IXP collects all flows
crossing its infrastructure\footnote{The IXP implements a uniform packet
sampling policy to reduce load on its routers.} and sends them to \sysname in a
{\em stream} (flows are sent out when they are closed by a FIN packet, in case
of TCP, or after a configurable timeout managed by the IXP operators).
\sysname mines this stream of traffic flows to detect DRDoS attacks in near real time.

Given our focus on DRDoS attacks, we keep only flows whose protocol is UDP and whose source
port is related to a service that is known to be vulnerable to be used for
attack amplification. The set of source port numbers and related UDP services we
use in our current configuration of IXmon is inspired by previous
work~\cite{AmplificationHell, USCERT-UDPAmp} and listed in Table~\ref{tab:udp_ports}.

\begin{table}[htbp]
\footnotesize
\caption{List of monitored UDP source ports}
\label{tab:udp_ports}
\begin{tabular}{c|c|c}

{Service} & {Port} 
                  & {\begin{tabular}[c]{@{}c@{}}Bandwidth\\ Amplification \\                        Factor\end{tabular}}                                   \\ \hline
                                                                                     \hline
DNS               & 53            & 28 to 54                                         \\ \hline
NTP               & 123           & 556.9                                            \\ \hline
CLDAP             & 389           & 56 to 70                                         \\ \hline
CharGen           & 19            & 358.8                                            \\ \hline
Memcached         & 11211         & 10,000 to 51,000                                 \\ \hline
SunRPC            & 111           & 7 to 28                                           \\ \hline 
SSDP              & 1900          & 30.8                                              \\ \hline 
SNMP              & 161           & 6.3                                              \\ \hline 
SRCDS             & 27005         & -                                          \\ \hline 
Call of Duty      & 20800         & -                                          \\ \hline 
NETBIOS           & 137           & 3.8                                               \\ \hline 
RIP               & 520           & 131.24                                           \\ \hline 
Quake             & 27960         & 63.9                                           \\ \hline 
Steam             & 29015         & 5.5                                          \\ \hline 
QOTD              & 17         & 140.3                                          \\ \hline   \hline

\end{tabular}
\end{table}

To analyze the continuous large stream of UDP flows received by \sysname, we proceed as follows.
First, \sysname partitions time into intervals of fixed length $\Delta{t}$ (one
minute, in our experiments). Given the set of all flows received during an
interval $\Delta{t}$, 
we map $srcIP$ and $dstIP$ to their respective AS numbers, $srcAS$ and $dstAS$
(e.g., using RouteViews data~\cite{routeview}). This gives us flows of the form: 
\begin{equation*}
    F_i(t)=(srcAS_i, srcPort_i, dstAS_i, dstPort_i, packets_i(t), bytes_i(t))
\end{equation*}
where $t$ indicates the start of a time interval $\Delta{t}$, $protocol$ is
omitted since it is constant (always UDP), and the packets and bytes counts vary
in time while the other flow parameters are fixed for a given subscript index.
Then, given a time interval $\Delta{t}$, we aggregate
all flows $F_i(t)$ that share the same source port and destination AS numbers,
and sum up all of their bytes. More formally, we obtain
aggregate {\em sketch} flows of this form:
\begin{equation*}
    A_k(t)=(srcPort_k, dstAS_k, bytes_k(t))
\end{equation*}
where $bytes_k(t)$ is the sum of the byte counts contributed by all flows
aggregated into $A_k(t)$.

Notice that, given a fixed pair of source port, $srcPort_k$, and destination AS,
$dstAS_k$, the related AS-level flows $A_k(t)$ give us a time series of the
total traffic volume (i.e., $bytes_k(t)$) flowing through the IXP that
originated from $srcPort_k$ (from any source IP) and destined towards $dstAS_k$
(to any
destination IP belonging to that AS and any destination UDP port). Also, while
not represented in the above {\em sketch}, for simplicity, we keep track of the
contribution (in terms of total bytes) to flow $A_k(t)$ of each
$srcAS_i$ whose traffic is aggregated into the sketch.

\subsection{Online Time Series Anomaly Detection}
\label{sec:timeseries}

Given a stream of flow sketches $A_k(t)$ related to a $(srcPort_k,
dstAS_k)$ pair, we detect anomalous increases in traffic volume by performing an
online analysis of the time series represented by $bytes_k(t)$. 

Specifically, we
maintain a time series model consisting of an exponentially-weighted moving average and variance~\cite{ewmav}, as follows:
\begin{equation}
    \mu(t) = \alpha \cdot \mu(t-1) + (1-\alpha) \cdot b(t)
    \label{eq:ewma}
\end{equation}
\begin{equation}
    \sigma^2(t) = (1-\alpha) \cdot (\sigma^2(t) + \alpha \cdot (b(t)-\mu(t-1))^2) 
    \label{eq:eq:ewmv}
\end{equation}
where $\alpha$ is a constant and where we omitted the subscript $k$ and used $b(t)$ in place of $bytes_k(t)$, for brevity.

Then, given the moving average, $\mu_k(t)$, and variance $\sigma_k^2(t)$
computed at time $t$ for $A_k(t)$, we compute an anomaly (or {\em deviation})
score as:
\begin{equation}
    \delta_k(t) =  \max \left( 0, \frac{b_k(t) - (\mu_k(t) + \theta \cdot \sigma_k(t))}{b_k(t) + \varepsilon} \right)
\label{eq:ddosdetect}
\end{equation}
where $\theta$ is a tunable parameter (set to 3 in our experiments) and
$\varepsilon$ is a small constant (e.g., $10^{-6}$) that
is only needed to avoid division by zero. Essentially, $\delta_k(t)$ tells us
how much $b_k(t)$ deviates (on the positive side) from the moving average plus a
tolerance factor proportional to the standard deviation. Notice that
$\delta_k(t) \in [0,1]$, which we use as an anomaly score. The larger
$\delta_k(t)$, the more strongly the current reading of $A_k$'s traffic volume,
$b_k(t)$, deviates from the expected value plus some tolerance that takes natural
variations into account. If $\delta_k(t) > \tau$, where $\tau$ is
a tunable detection threshold (set to 0.5 in our experiments), we say that the
current reading of the traffic volume for the flows aggregated by $A_k$ is anomalous.

\vspace{5pt}
\noindent
{\em Additional details}: At every new time interval, we use
Equations~\ref{eq:ewma} and ~\ref{eq:eq:ewmv} to update our time series
model. However, once an anomaly is detected, we stop updating the model until
the new traffic volume measurements go back to pre-anomaly levels. More
formally, assume $t_d$ is the first time in which an anomaly is detected,
we do {\em not} use the new measurement at time $t_d$ to compute $\mu(t_d+1)$ and 
$\sigma^2(t_d+1)$. Now, let 
\begin{equation}
    \delta_k(t+n, t) =  \max \left( 0, \frac{b_k(t+n) - (\mu_k(t) + \theta \cdot \sigma_k(t))}{b_k(t+n) + \varepsilon} \right)
\label{eq:ddosdetect2}
\end{equation}
and $t_d = t+1$. In other words, at the time when the anomaly is detected,
$n=1$. At the next time slot, $n=2$, we compare the latest measurement of the
traffic volume $b_k(t+n)$ to the time series model that was last updated at time
$t$. If $\delta_k(t+n, t) > \tau$ this means that the anomalous traffic is still
present at time $t+n$, and we continue to keep the same model computed at
time $t$. Let us now assume that at $n=m$ the anomalous levels of traffic
revert back to normal. Namely, $\delta_k(t+m, t) \le \tau$. Then, we use
$b_k(t+m)$ to update the values of $\mu_k$ and $\sigma_k$ and keep updating the
model at the following time intervals, until another anomaly is identified.

This approach of updating the average and standard deviation only during
``normal times'' allows us to more easily determine when a traffic volume
anomaly, which may represent a DRDoS attack, starts and ends. Specifically, in
the example above we can determine that the anomaly started at time $t+1$ and
ended at time $t+m$.

\subsection{Attack Detection}
\label{sec:detection}

Let now $A_k(t)$ be a traffic sketch time series, and assume that $t_d$ is the
time interval in which a time series anomaly has been detected using the
approach described in Section~\ref{sec:timeseries}. To detect DRDoS attacks and
filter out possible traffic volume anomalies unrelated to reflection attacks, we
introduce two additional conditions:
\begin{itemize}
    \item {\em Minimum traffic volume}: Given the last aggregate traffic volume
    measurement, $b_k(t_d)$, we discard the detected anomaly if $b_k(t_d) <
    \nu$ (in our experiments we set $\nu$ to 5Mbps). The reason is that if the aggregate traffic volume is very low, either
    the anomaly is not caused by an attack, or the effects of the attack on the
    target AS's bandwidth are negligible and can be ignored.
    \item {\em Source AS volume entropy}: Since we focus on DRDoS attacks, we
    expect the anomalous traffic volume increase to be distributed across
    multiple reflectors located in different source ASes.
\end{itemize}

To compute the source AS volume entropy, we first consider the set of source ASes whose traffic is
aggregated into $A_k$, and take into account the overall number of bytes sent
from each of this sources ASes to $A_k$'s destination AS (i.e., the
potential victim network). Let $S_k(t_d) = \{s_1, s_2, \dots, s_n\}$ represents the
set of traffic volume amounts contributed by each source AS at time $t_d$. We
then normalize each element in the set as $s'_i = \frac{s_i}{\sum_{j=1}^{n}
s_j}$. Finally, we treat $s'_i$ as the probability of ``observing'' the $i$-th
source AS as contributor to $A_k$'s aggregate traffic, and compute the entropy
$\mathcal{H}(S'_k(t_d))$ of the set $S'_k(t_d) = \{s'_1, s'_2, \dots, s'_n\}$.
If $\mathcal{H}(S'_k(t_d)) = 1$, it means that the traffic from port $srcPort_k$
to $dstAS_k$ is evenly distributed across the contributing source ASes. On
the other hand, low values of $\mathcal{H}(S'_k(t_d))$ mean that most of the
traffic is contributed by only one (or very few) source ASes. Therefore, we set
a threshold $h$ so that traffic volume anomalies are labeled as DRDoS attacks
only when $\mathcal{H}(S'_k(t_d)) > h$ (in our experiments, we use $h=0.4$).

All time series anomalies detected based on the algorithm described in
Section~\ref{sec:timeseries} that also meet the two above conditions are labeled
as DRDoS attacks. Correspondingly, a DRDoS attack alert is issued, which
contains all details of the attack as measured at time $t_d$, including the
destination AS number, source port, 
current aggregate attack volume, and distribution of traffic amounts from the
contributing source ASes. A new alert is issued for every new time interval
$t_d+n$ for which the attack is sustained, allowing a network operator to
identify whether the attack is still ongoing or has terminated (when no new
alert is issued). On the other hand, time series anomalies that do not pass the
checks discussed above are logged and can be sent to network operators but are not labeled as DRDoS attacks.

\section{Analysis of In-the-Wild Attacks}
\label{sec:measurements}

In this section we provide some background information about SoX, describe how
we setup and deployed \sysname at SoX, and present our measurements and analysis
of the in-the-wild DRDoS attacks we detected during our deployment period.
Notice that in the following analysis we anonymize all AS numbers related to
autonomous systems involved in the detected DRDoS attacks, as some of this
information may be sensitive (e.g., some of SoX's members may not want to
publicly disclose how many attacks their network received and if/how they
mitigated them). For instance, we replace AS 10490 with a consistent but
randomly chosen identifier of the form ``Anon.XXX'' (where XXX is a positive
integer). 

\begin{figure*}[!ht]
    \centering
    \includegraphics[scale=0.6]{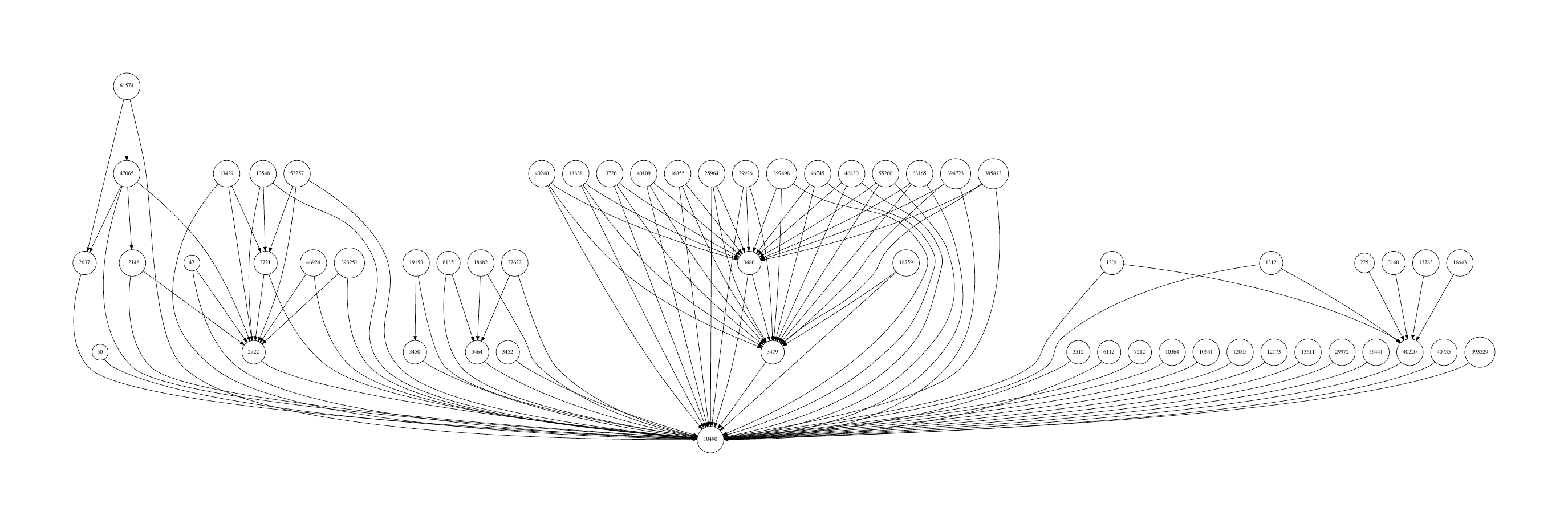}
    \caption{Networks reachable from the Internet through SoX (AS 10490). Directed edges represent a customer->provider relationship.}
    \label{fig:sox_reachable}
\end{figure*}

\subsection{\sysname Implementation and Setup}
\label{sec:setup}

We implemented \sysname's flow parsing and traffic aggregation modules in C++,
leveraging an open-source tools named FastNetMon~\cite{FastNetMon}. FastNetMon
is a DDoS detection system mainly geared towards enterprise networks.
Unfortunately, we found that the attack detection approach used by FastNetMon
was not designed to detect and track DRDoS attacks related to many possible
large networks and involving large numbers of source ASes, making it completely
unusable in our deployment scenario. Therefore, while we leveraged and adapted
the NetFlow parsing module of FastNetMon, we designed and implemented our own
IXP-focused online time series anomaly detection and DRDoS detection algorithms
using Python (we plan to share \sysname's code on GitHub after publication).

As explained in Section~\ref{sec:ixmon}, \sysname's online anomaly detection and
DRDoS detection algorithms include a few tunable parameters. It is worth noting
that our main objective for developing \sysname was to enable the study of
in-the-wild attacks. Because sizable ground truth datasets of in-the-wild DRDoS
attacks are very difficult to come by (we are not aware of any publicly
available dataset of this kind), to tune the parameters we rely on domain
knowledge and a manual analysis of \sysname's logs during the preliminary phases
of our deployment. In general, we take a conservative approach that favors
minimizing possible false detections. While this may cause us to miss some
smaller (i.e., lower volume and duration), more subtle DRDoS attacks, these
attacks are unlikely to have a large impact on their target networks.

We set the length of the time interval for traffic aggregation $\Delta{t} = 1$
minute. This interval is long enough to accumulate sufficient aggregate data
from the stream of flows related to each $(srcPort_k, dstAS_k)$ pair and to
compute meaningful traffic sketches, and at the same time it enables near
real-time DRDoS detection. 

In Equation~\ref{eq:ddosdetect}, we set the parameter $\theta=3$. Essentially,
$\theta$ controls how much the traffic volume can deviate from the mean, before
an anomaly is detected. The value of $\theta=3$ is quite conservative, and is
inspired by the fact that for Gaussian distributions $Pr(\mu-3\sigma \le X \le
\mu+3\sigma) \approx 99.73\%$. In addition, we set the anomaly detection
threshold $\tau=0.5$. In other words, we tune the system to detect large
anomalies, as compared to historic traffic volumes modeled by the moving average
and standard deviations. While this may cause us to miss small (i.e., low
volume) attacks, it makes sure that the anomalies we detect are in fact highly
likely related to attacks. This is further reinforced by the additional
constraints explained in Section~\ref{sec:detection}. 

As for the parameters defined in Section~\ref{sec:detection}, we set $\nu=5$Mbps
because DRDoS attacks whose peak traffic is lower are unlikely to cause much
disruption to institutional networks (such networks typically have Internet
connectivity bandwidth ranging from hundreds of Mbps to tens of Gbps). Finally
we set the source AS entropy threshold $h=0.4$. We tuned this threshold based on
a data collected during a preliminary deployment of \sysname, and is meant to
capture attacks whose traffic is fairly distributed across multiple sources,
rather than all coming mostly from one single source AS. 

An additional ``operational'' parameter is related to the packet sampling rate
used by the network operator that provides the raw flows. In \sysname, we take
the sampling rate into account, and adjust our traffic measurements accordingly
(e.g., we adjust the average traffic volume measured per minute of observation).

\subsection{Data Collection at SoX}
\label{sec:data_collection}

As mentioned earlier, we deployed IXmon at a large IXP called SoX (AS 10490)
that provides peering and Internet connectivity services to several research and
education networks. Specifically, based on public data provided on AS-to-AS
relationships provided by CAIDA~\cite{CustomerCones, CAIDA_cones}, SoX has more
than 20 direct customer networks (also called the IXP {\em members} or {\em
participants}), peers with 9 other large ASes, and is connected to 5 upstream
providers, as shown in Figure~\ref{fig:sox_direct_links} (in appendix). Furthermore, SoX
serves as upstream provider for a variety of smaller ASes that are reachable
through it from the rest of the Internet, as shown in
Figure~\ref{fig:sox_reachable}. While our deployment is still ongoing, this
study is based on data collected between {\em April,2018 - April,2020} (due to interruptions due to
operational reasons, our traffic monitoring was only active during part of this
time period). Overall, we collected traffic information for 634 days. During
this period, the source/destination traffic crossing the IXP's fiber was related
to a total of {\em 5212} different autonomous systems.

\begin{figure*}[!ht]
    \centering
       \begin{subfigure}{0.33\textwidth}
           \centering
           \includegraphics[scale=0.37]{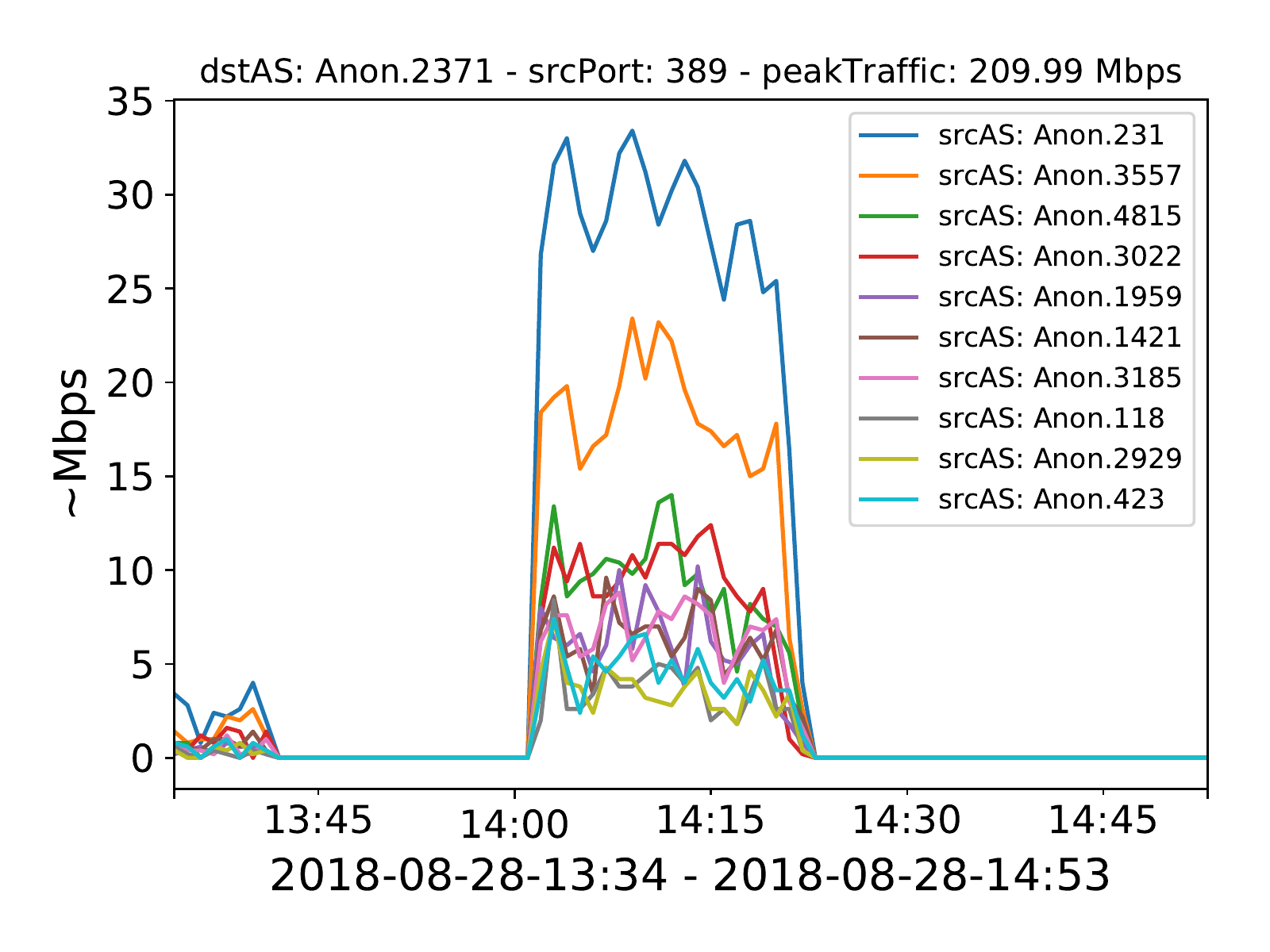}
           \caption{CLDAP-based attack}
           \label{fig:suspicious_ads_1}
       \end{subfigure}%
       \begin{subfigure}{0.33\textwidth}
           \centering
           \includegraphics[scale=0.37]{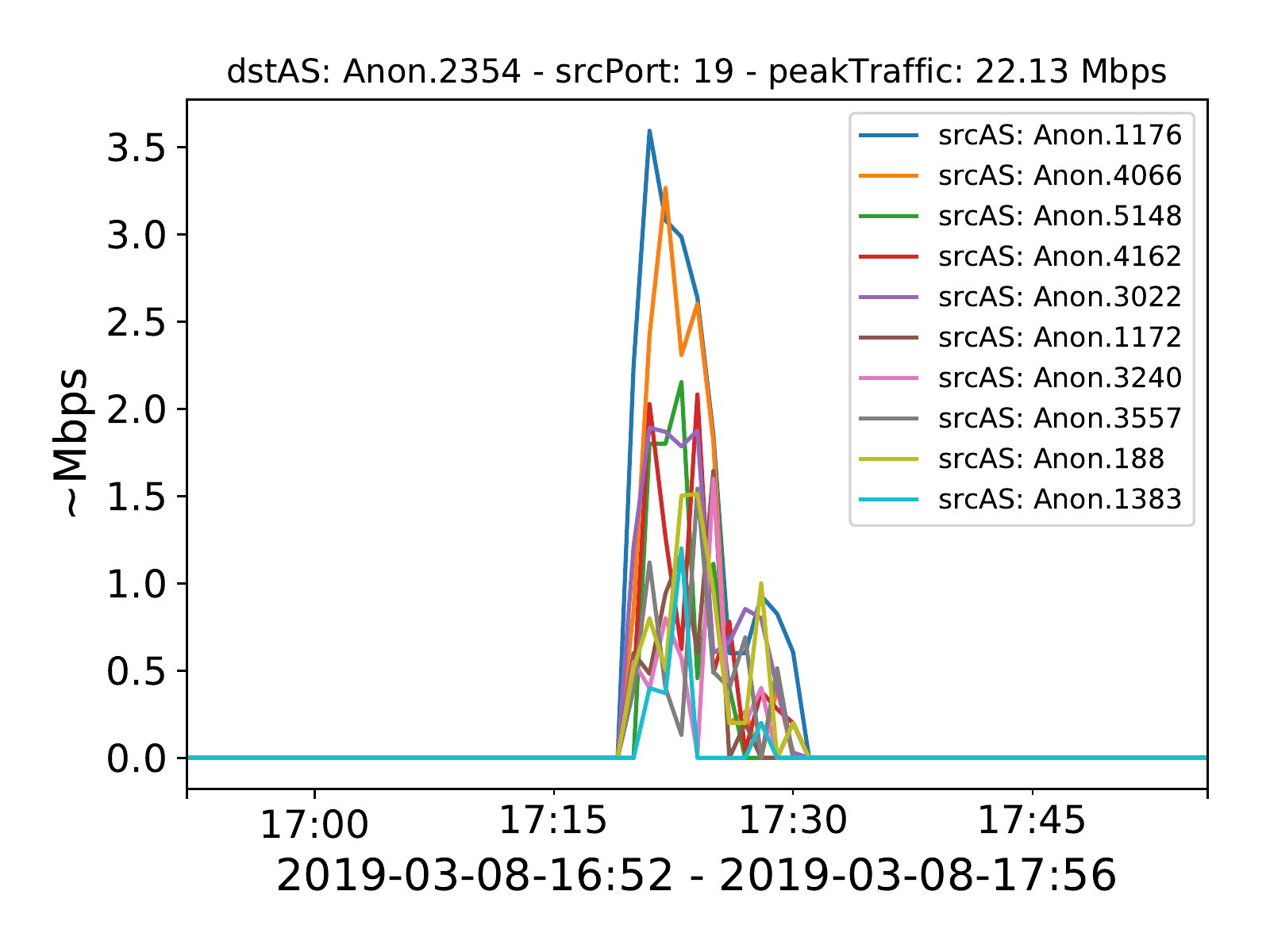}
           \caption{CHARGEN-based attack}
           \label{fig:suspicious_ads_2}
       \end{subfigure}%
       \begin{subfigure}{0.33\textwidth}
           \centering
           \includegraphics[scale=0.37]{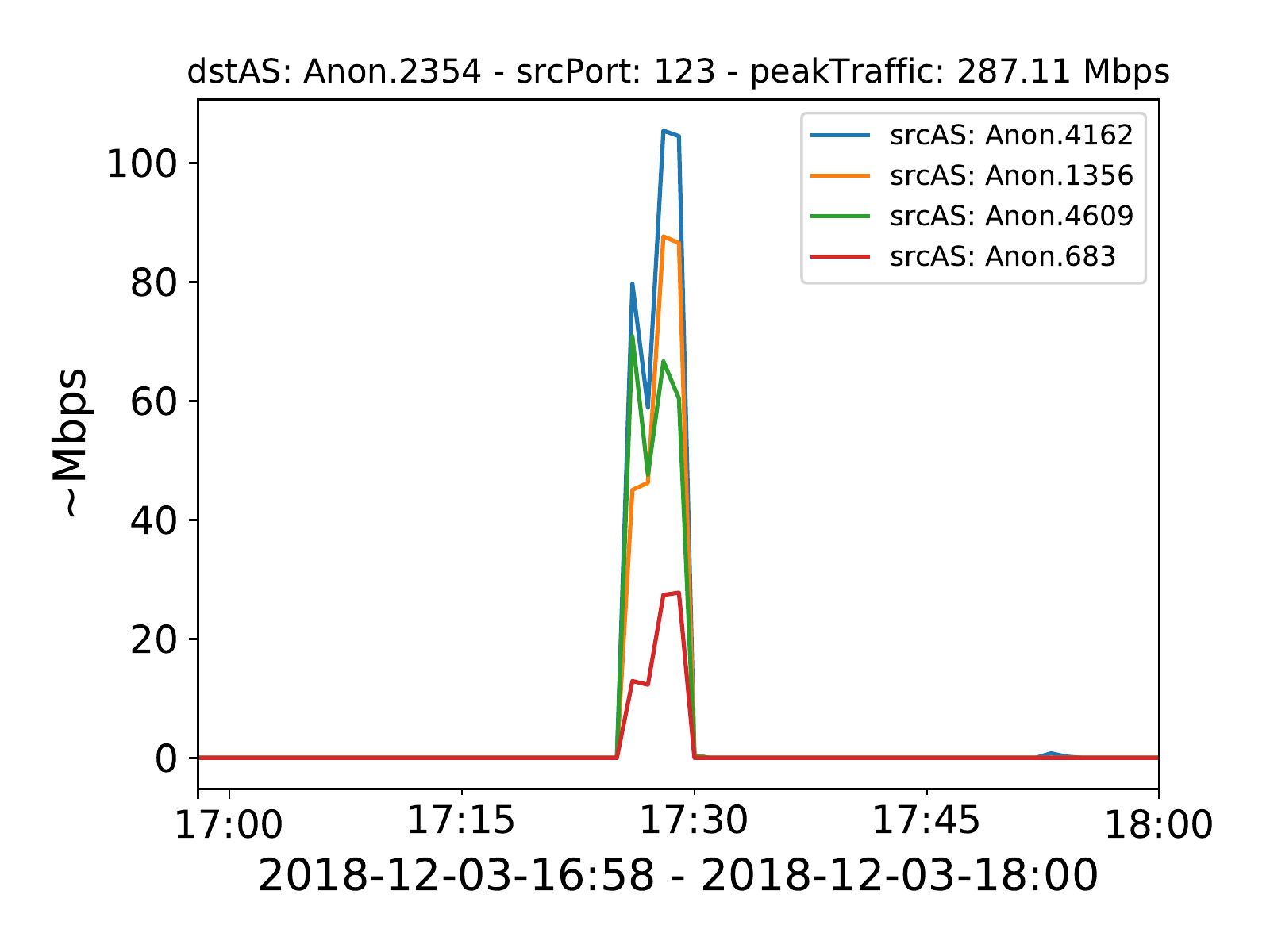}
           \caption{NTP-based attack}
       \end{subfigure}
       \caption{Examples of DrDoS attacks detected by \sysname}
       \label{fig:drdos_attacks}
\end{figure*}

\subsection{Attack Measurements and Analysis}
\label{sec:analysis}

In this section, we present a detailed analysis of the DRDoS attacks detected by
\sysname to understand their behavior and gain insights that could prove useful
for mitigating future attacks.

As an example of the attacks that are included in the analysis provided below,
Figure~\ref{fig:drdos_attacks} shows a snapshot of three different DRDoS attacks
detected by \sysname. The $x$ axis show that time window within which the attack
occurred (including a duration of 30 minutes prior to and after the attack), whereas the $y$ axis shows the volume of traffic contributed by each
source AS involved in the attack (the graph is limited to the top 10 source ASes
by volume). Each line in the graph represents the traffic sent to the victim AS
from a single source AS. For instance, Figure~\ref{fig:suspicious_ads_1} shows a DRDoS attack
that leverages the CLDAP service (source port 389) directed towards AS
Anon.2371. The aggregate traffic for the attack, which sums the contribution of
all source ASes that sent traffic to AS Anon.2371 from UDP port 389 reached a peak
of $\approx$210Mbps. It is interesting to notice that before and after the
attack there was little or no traffic sent by those source ASes to the
destination AS from port 389. Then, all of a sudden all the source ASes start
sending high volumes of traffic in a coordinated way, which is a telltale sign
of an ongoing DRDoS attack. After all, inter-AS CLDAP use is rarely needed
or justified, and it is therefore natural to have very low or no inter-AS CLDAP
traffic outside of DRDoS attacks. In addition, having many source ASes sending
CLDAP traffic to a common destination AS would be quite a big coincidence for
this to be explained by normal activities.

\begin{figure}[!ht]
 \centering
    \begin{subfigure}{\columnwidth}
        \centering
        \includegraphics[scale=0.4]{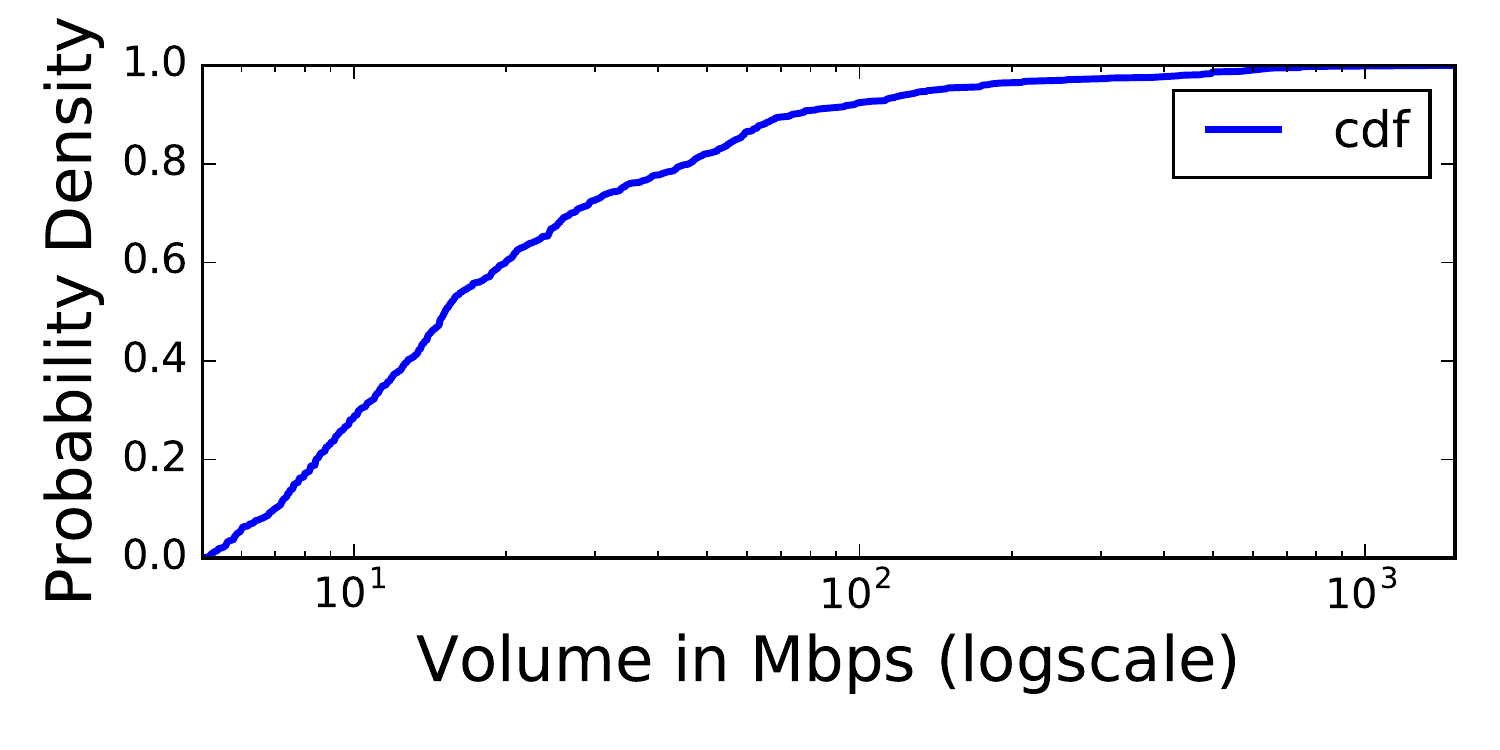}
        \caption{Peak attack volume}
        \label{fig:vol_cdf}
    \end{subfigure}
 
    \begin{subfigure}{\columnwidth}
        \centering
        \includegraphics[scale=0.4]{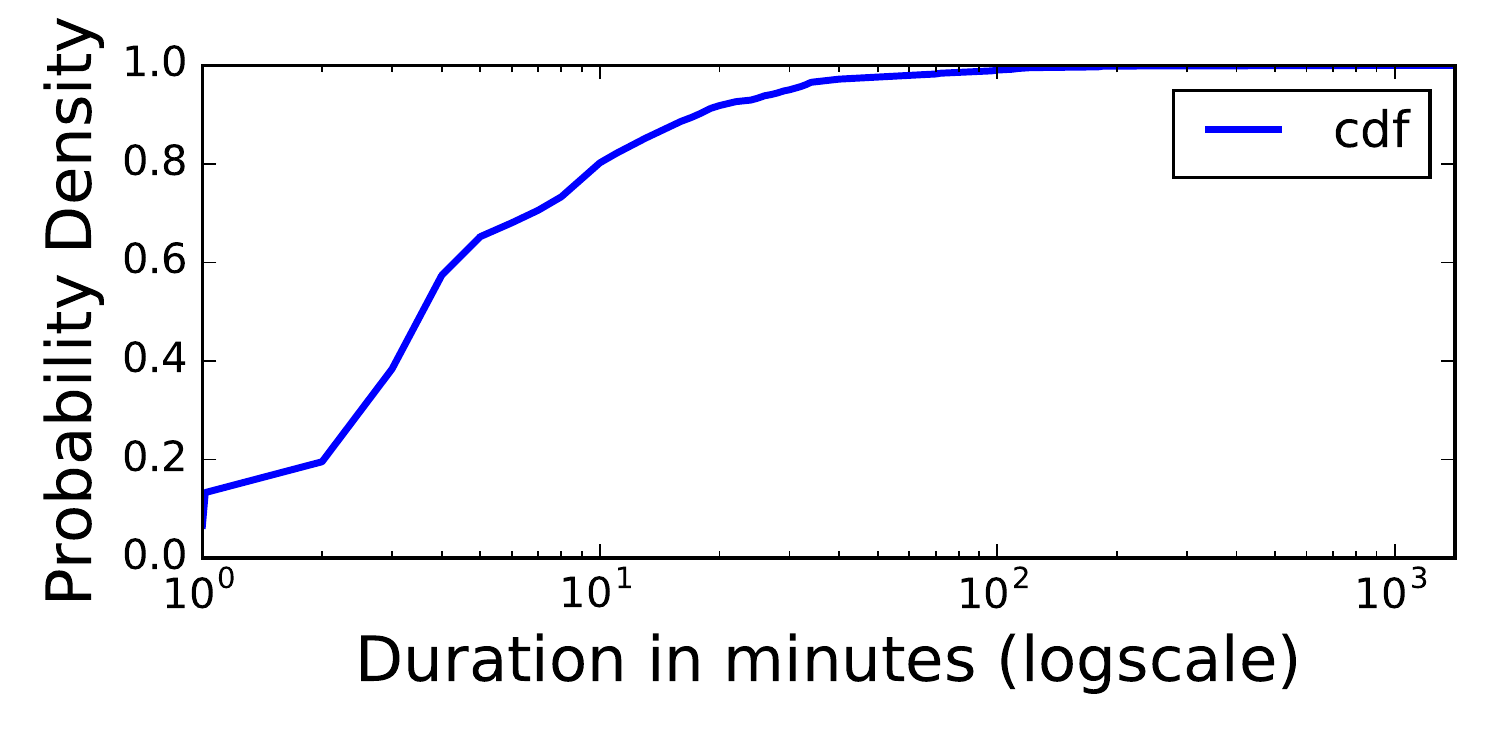}
        \caption{Attack duration}
        \label{fig:time_cdf}
    \end{subfigure}
    \caption{Distribution of peak attack volumes and durations}
\end{figure}

\begin{figure}[!ht]
\begin{center}
\includegraphics[scale=0.3]{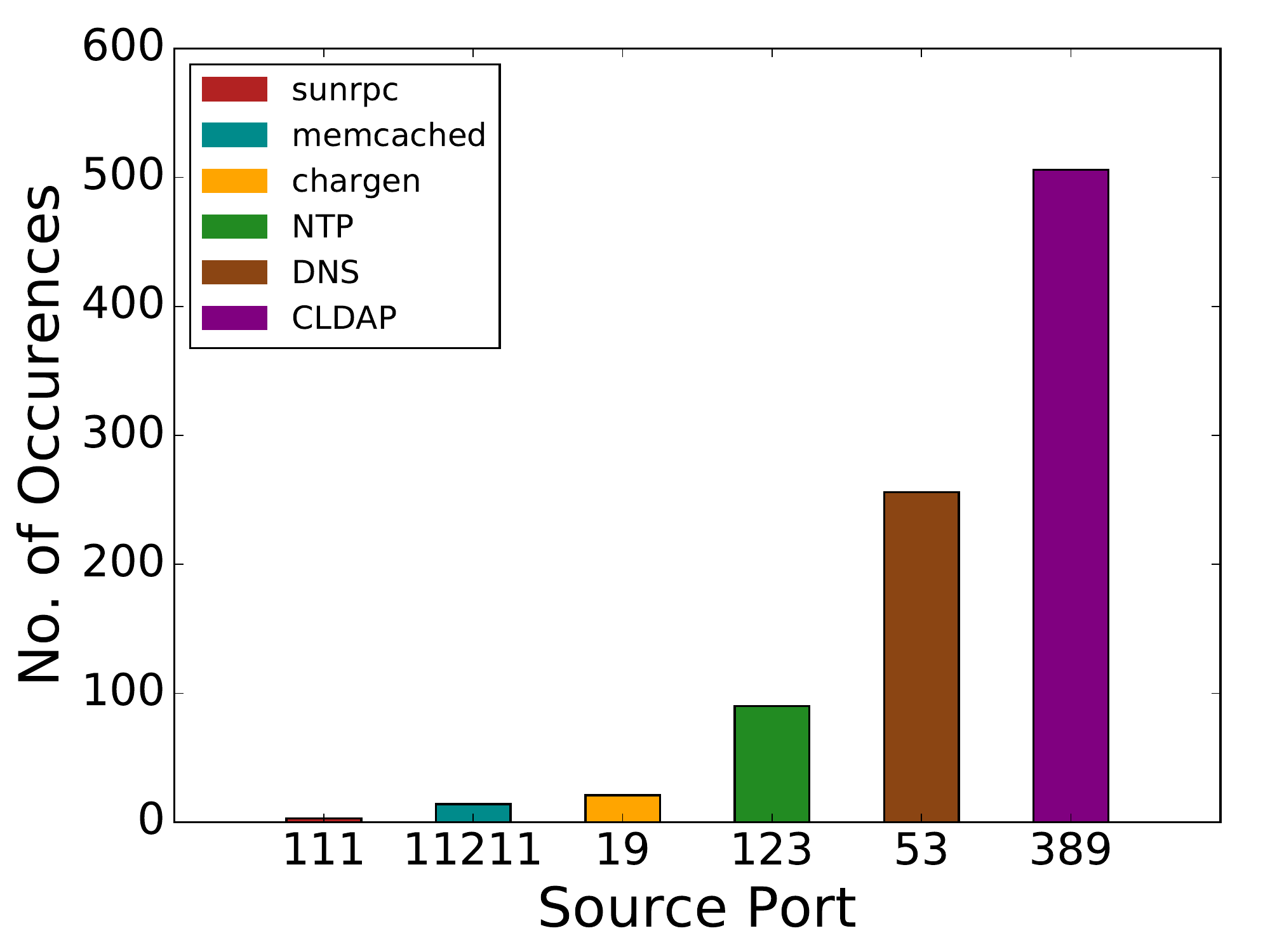}
\caption{Number of attack instances per (reflection) source port}
\label{fig:ports}
\end{center}
\end{figure}

\subsubsection{Volume and Duration}
\label{sec:voldur}

While large DDoS attacks have caught the attention of bloggers and news media,
there is limited publicly available data on the overall distribution and
characteristics of in-the-wild DRDoS attacks (some information can be found in a
2017 blog post by Cloudflare~\cite{Cloudflare-reflections}).
%

To better understand in-the-wild DRDoS attacks, we analyze the
characteristics of all attacks detected by IXmon.
Specifically, during our deployment period \sysname detected 987 attacks. We use this large
number of attacks to measure the distribution of the volume and duration of
in-the-wild DRDoS attacks, which are reported in Figures~\ref{fig:vol_cdf}
and~\ref{fig:time_cdf}. It can be seen that most of the observed attacks
($\approx 80\%$) have a duration of less than 10 minutes, whereas the median
peak attack volume is less than 20Mbps. Overall, only {\em $\approx 8\%$} of the
attacks reach a peak volume of more than 100Mbps with a few attacks reaching
peaks above 1Gbps (the highest attack volume we observed was {\em 1.5Gbps}).

\subsubsection{Services Abused for Attack Amplification}
\label{sec:ports}

\begin{figure}[!ht]
\begin{center}
\includegraphics[scale=0.33]{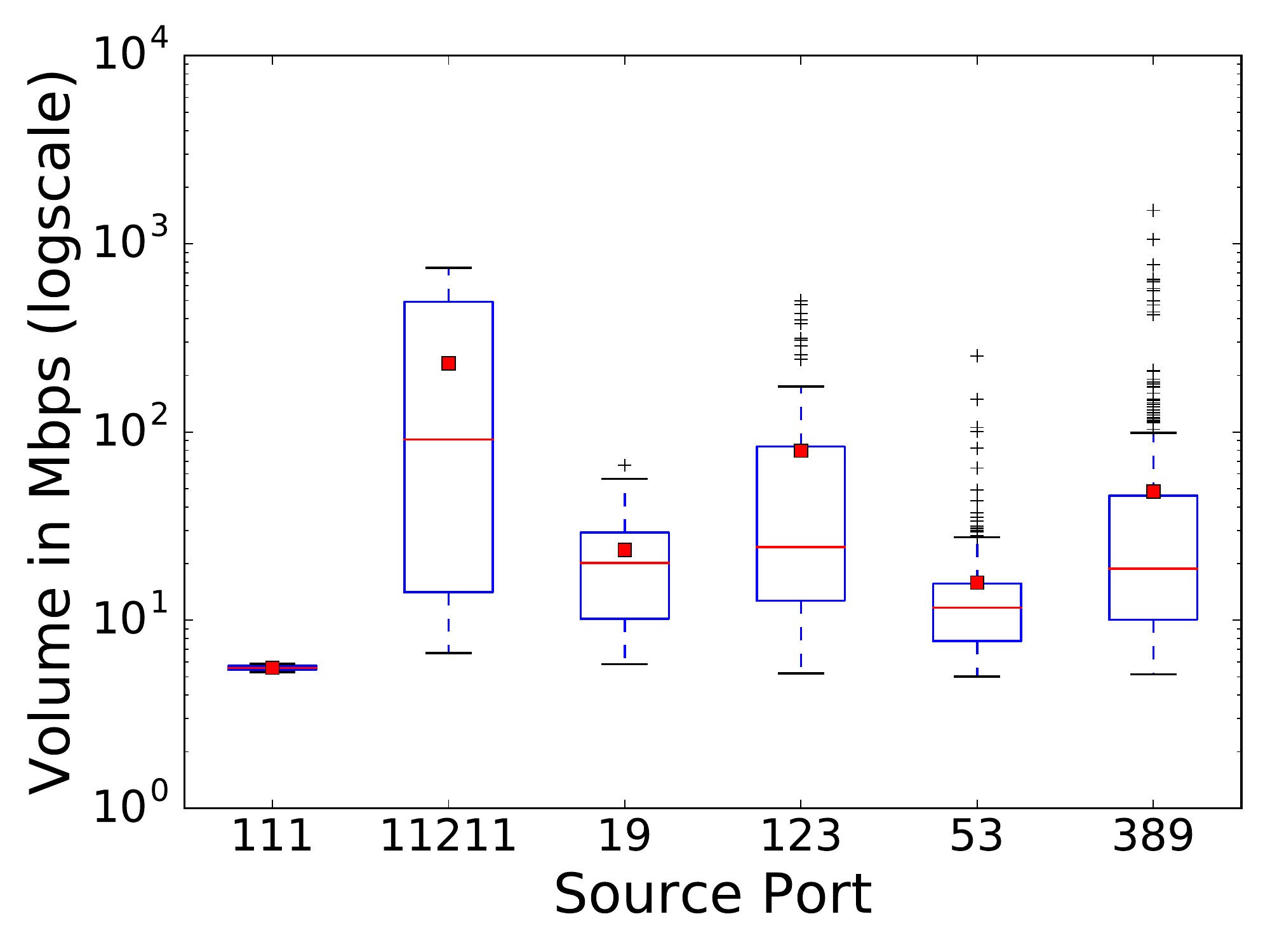}
\caption{Distribution of attack volume per (reflection) source port}
\label{fig:ports_vol}
\end{center}
\end{figure}

\sysname monitors traffic from the UDP ports listed in
Table~\ref{tab:udp_ports}. However, only some of these ports were used in DRDoS
attacks visible from SoX. Figure~\ref{fig:ports} shows the distribution of
source ports (ab)used for reflecting traffic against DRDoS victims, with the $y$
axis showing the number of attacks in which a given port was used. As can be
seen, CLDAP (port 389) appears to be the most abused service for attack
amplification, followed by DNS (port 53) and NTP (port 123). 
Figure~\ref{fig:ports_vol} reports a boxplot showing the distribution of peak
attack volume per port (the red line represents the median, while the red square
shows the average value). This shows that some CLDAP-based attacks reached
peak volumes above 1Gbps. 

\begin{figure}[!ht]
\includegraphics[width=0.8\linewidth]{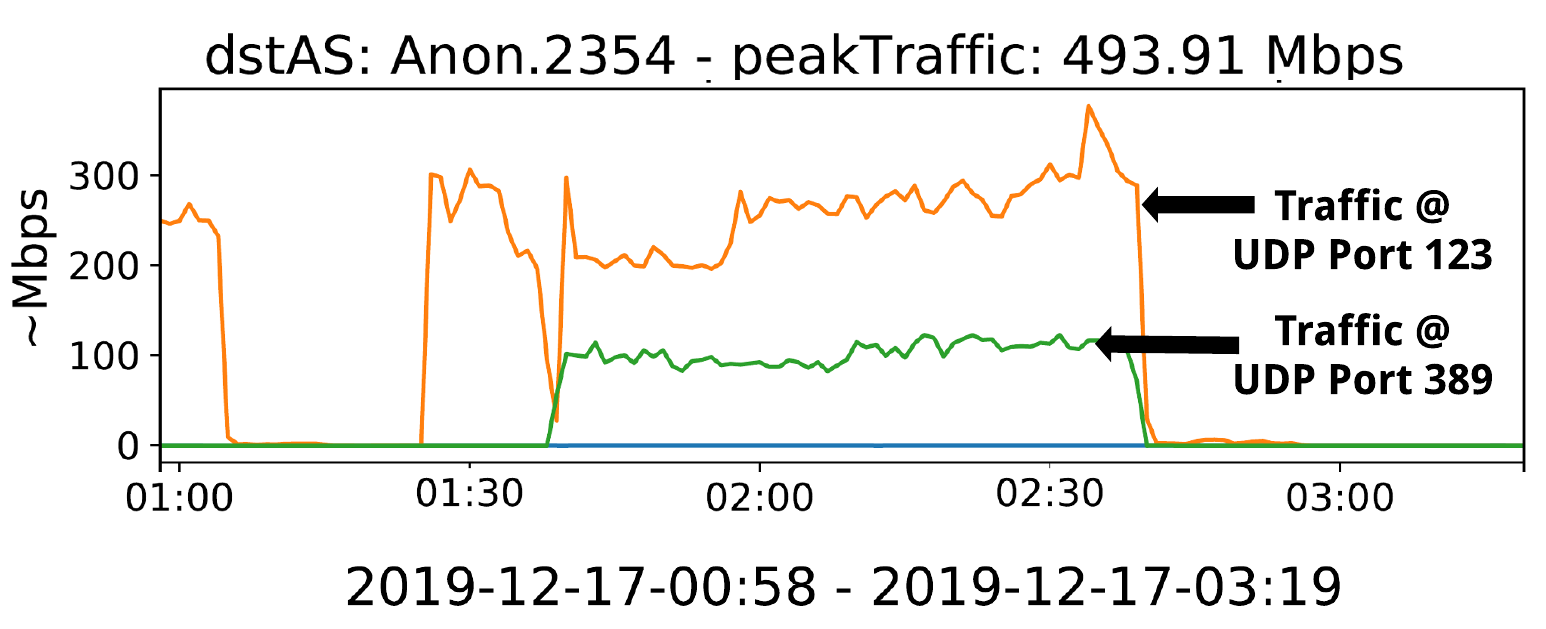}
\caption{Example of a multi-vector attack 
}
\label{fig:multivector_attack}
\end{figure}

\subsubsection{Multi-Vector Attacks}
\label{sec:multivector}

DRDoS attacks can be launched by abusing more than one UDP service at a time.
Currently, \sysname separately tracks traffic from a given source port and
detects DRDoS attacks independently for each abused service. However, attackers
can abuse multiple services at the same to increase the number of reflectors to
be aimed against the victim and thus further amplify the attack bandwidth. To
analyze these attacks in our alerts dataset, we can easily retrieve DRDoS
attacks related to individual source ports that have a common destination AS and
that overlap in time. By doing so, we found 36 multi-vector attack instances
(out of 987) involving up to 4 different source ports simultaneously. 
Figure~\ref{fig:multivector_attack} shown an example of attack detected using
\sysname that simultaneously leverages NTP (port 123) and CLDAP (port 389) to
reflect the attack traffic towards AS Anon.2354. A coordinated surge in traffic
volume can be observed from both source ports, clearly indicating a multi-vector
attack.

\subsubsection{Distribution of Reflectors}
\label{sec:source_distribution}

DRDoS attacks are executed by exploiting a (at times large) number of publicly
reachable reflection servers. In this section, we analyze where reflected attack
traffic originates from.

\begin{figure}[!ht]
\begin{center}
\includegraphics[scale=0.4]{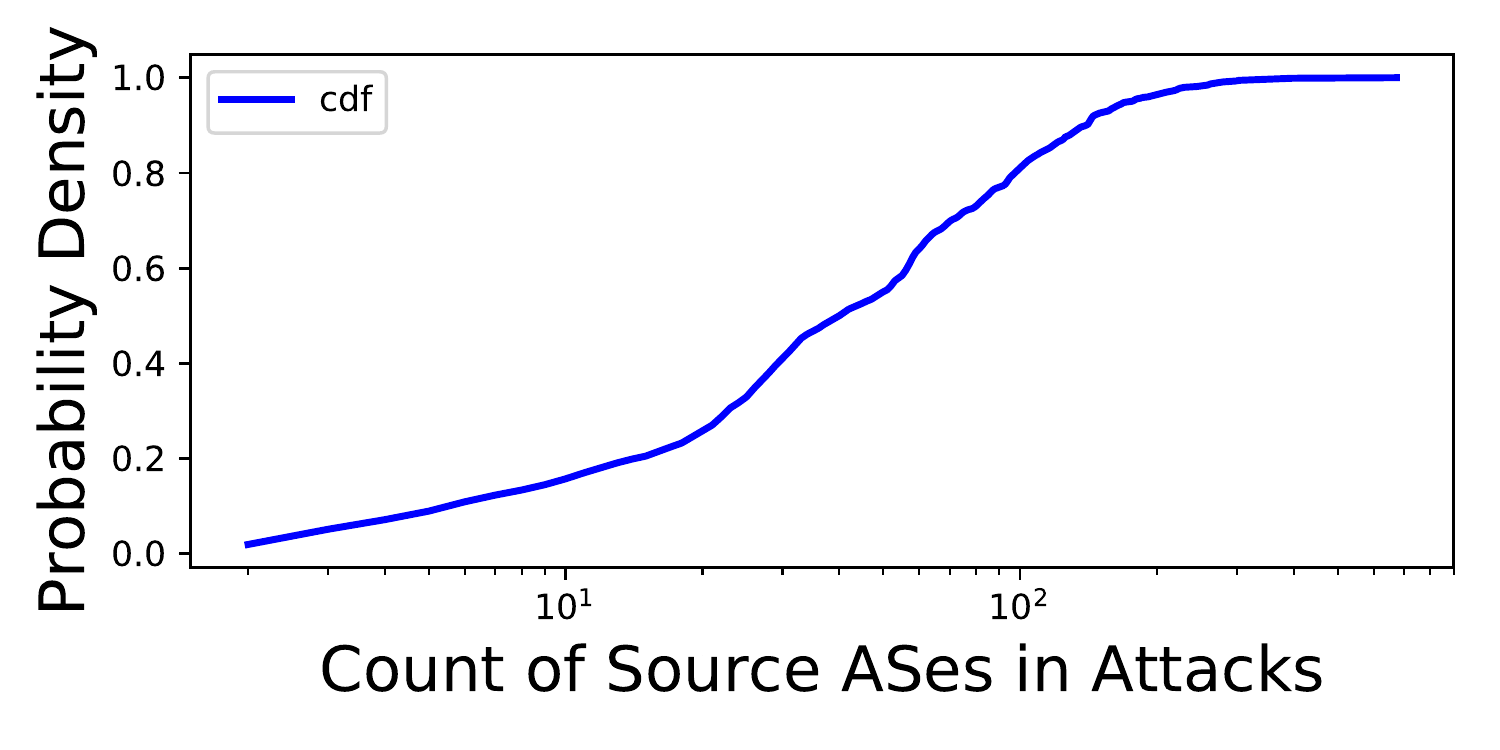}
\caption{Distribution of number of source ASes involved in attacks}
\label{fig:sources_count}
\end{center}
\end{figure}
\begin{figure}[!ht]
\begin{center}
\includegraphics[scale=0.4]{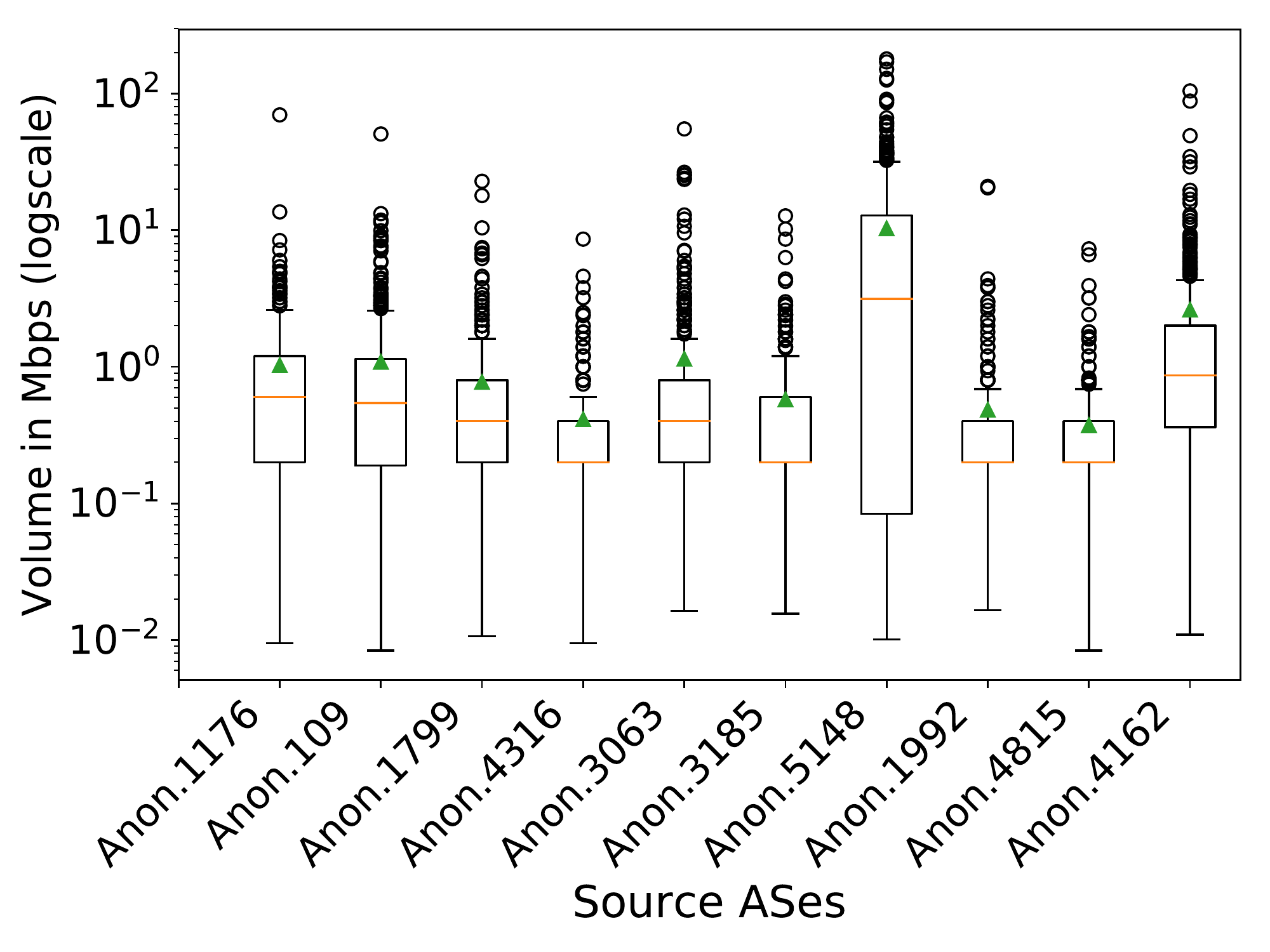}
\caption{Distribution of peak traffic volume contributed by the top 10 Source ASes in Attacks}
\label{fig:attack_sources_top_10}
\end{center}
\end{figure}

Figure~\ref{fig:sources_count} shows the distribution of the number of different
source ASes that contribute to each attack. The median is 40, indicating that at
least half of all attacks are highly distributed across many different source
networks that are themselves abused to reflect and amplify attack traffic. In
Figure~\ref{fig:attack_sources_top_10} we show the distribution of peak traffic
contributed to different attacks by the top 10 source ASes (ranked based on the
number of DRDoS attacks each source AS participates to). As can be seen, the
median (red line) peak volume for reflected traffic from each AS is relatively
limited, typically around $\approx 1$Mbps, though there are also significant
outliers with high peak traffic volumes. Either way, when combining together all
contributing source ASes the
attacks these ASes facilitate can easily reach hundreds of Mbps.

\begin{figure}[!ht]
\begin{center}
\includegraphics[scale=0.6]{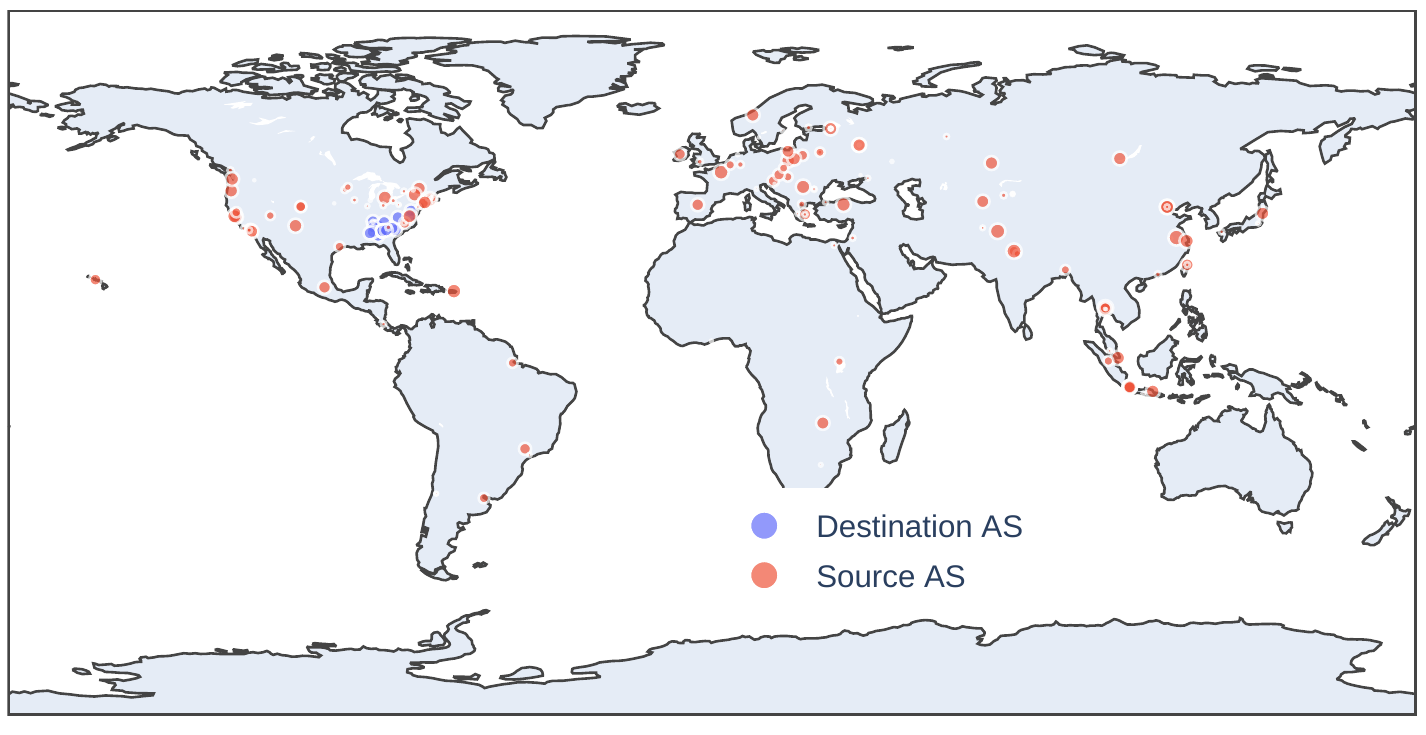}
\caption{Geo-locations of source and destination ASes for NTP-based DRDoS attacks}
\label{fig:ntp_geo}
\end{center}
\end{figure}

To analyze the geographical distribution of the networks where reflection
servers reside, we plot the location of the source ASes that contributed to the
DRDoS attacks detected by \sysname. To map the geolocation of a given AS we
first obtain the prefixes owned by the AS, based on BGP traffic from the day
before the AS participated in an attack. Next, we select a random IP address
belonging to one of the prefixes and map the IP address to its geolocation via
a IP geolocation API~\cite{ip_loc_api}. While this is only an approximate
method for determining the geolocation of an AS (some AS numbers span multiple
regions), it serves the purpose of giving an idea of how geographically
distributed the reflectors typically are. As an example,
Figure~\ref{fig:ntp_geo} shows the geolocation of both destination ASes (i.e.,
the victims) and source ASes (i.e., the networks that host the servers abused
for attack reflection and amplification) related to NTP-based attacks detected
by \sysname. Naturally, given the fact that SoX serves as a peering and
connectivity hub for research and education networks in the South-East USA, the
destination ASes are clustered in that region. It is easy to see that the
sources of NTP traffic are distributed widely across the world. This is
evidently anomalous, in that in normal (i.e., non-attack) cases the vast
majority of NTP responses would be coming from NTP servers that are
geographically closer to the the requesting IP address. Combined with the fact
that no NTP requests are sent from a victim AS to those reflection servers, this
lack of ``locality'' could be used as a way to develop an attack mitigation
strategy (see Section~\ref{sec:mitigation}).

\subsection{Attack Mitigation}
\label{sec:mitigation}

In this section, we analyze whether the operators of the victim networks
attempted to mitigate the attacks that were detected by \sysname. Specifically,
we focus on mitigations that require BGP actions (explained below).
Afterwards, we discuss how \sysname could help mitigate future attacks by (a)
detecting DRDoS attacks in near real time (with a delay of about $\Delta{t}=1$
minute); (b) determining the AS being targeted by the attack and what service
(i.e., source UDP port) is being abused to reflect/amplify attack traffic; and
(b) identifying the source ASes that contribute the most to the attack, so that
attack traffic originating from those ASes can be filtered out.

\subsubsection{Mitigation Strategies}: Multiple ways exist to respond to
DDoS attacks~\cite{drdos_survey}. However, as we focus on
bandwidth exhaustion DRDoS attacks, we ignore mitigations implemented locally at
the victim network. Instead, we focus on mitigations that are implemented
upstream, with the help of third-party networks such as traffic providers or
scrubbing centers, and that make use of BGP to drop or redirect traffic before
it reaches the victim network:
\begin{itemize}
    \item {\em Blackholing}: BGP-based blackholing redirects all traffic towards
    a victim AS (both legitimate and malicious traffic) into a null interface,
    or ``blackhole.'' Although multiple variations of blackholing exist, they
    are primarily achieved by adjusting the next-hop attribute and BGP
    communities in BGP announcements~\cite{BlackholingIETF}. The next-hop method
    involves the trigger source sending a BGP update to the edge routers with
    the next-hop attribute set to an IP address that is pre-configured to a {\em
    null} interface. The most commonly used next-hop IP for blackholing is
    192.0.2.1, which is reserved by IANA for test
    networks~\cite{IANA_reserved}.
    \item {\em Traffic re-routing}: In this method, all traffic towards the
    victim network is redirected to third-party services, such as a traffic
    scrubbing center that is capable of detecting and dropping DDoS attack
    traffic. Then, legitimate traffic is forwarded back to the original
    destination (i.e., the victim AS). To re-route traffic, a BGP announcement
    can be issued by the scrubbing center AS taking ownership of the victim's
    targeted IP prefixes, essentially performing an {\em authorized} BGP
    hijacking. After these BGP announcements propagate, all traffic destined to
    the victim AS will instead reach the scrubbing center. After the attack has
    ended, another BGP messages can be issued to reinstate the original IP
    prefix ownership and again route all traffic to the true destination.
\end{itemize}

\subsubsection{Detecting BGP-based DRDoS Mitigations}
To detect whether mitigations were put into place to counter the attacks
detected by \sysname, we perform an analysis of BGP announcements related to the
victim ASes before, during and after a DRDoS attack occurrence. To this end, we
leverage routing information from RouteViews~\cite{routeview}, as explained below:
\begin{itemize}
    \item {\em Blackholing}: To detect the use of blackholing mitigations, we
    monitor the BGP updates involving all IP prefixes owned by a victim AS, and
    check if any of these updates announce the next-hop to be 192.0.2.1. In
    addition, we look for BGP updates with a community value set to 666, which
    is commonly used to implement balckholing~\cite{BlackholingIETF}.  
    \item {\em Traffic re-routing}: To detect cases in which traffic is re-routed
    to a third-party AS (e.g., to a scrubbing center), we gather all BGP updates
    made around a DRDoS attack time window and consider all updates related to
    IPs that fall within the victim's network ownership. Then, for each such BGP
    update, we check if the origin AS (extracted from RIB records) has changed,
    compared to before the attack (e.g., compared to the previous day) . If the
    origin AS in the BGP updates observed during the attack does not match the
    previously seen origin AS, we mark this as a temporary change in ownership,
    and check whether future BGP messages also show another change of AS
    ownership from the third-party AS back to the previous origin AS. We
    implement this approach using \texttt{PyBGPStream} library and Routeview
    data. 
\end{itemize}

\begin{figure}[!ht]
    \includegraphics[width=\linewidth]{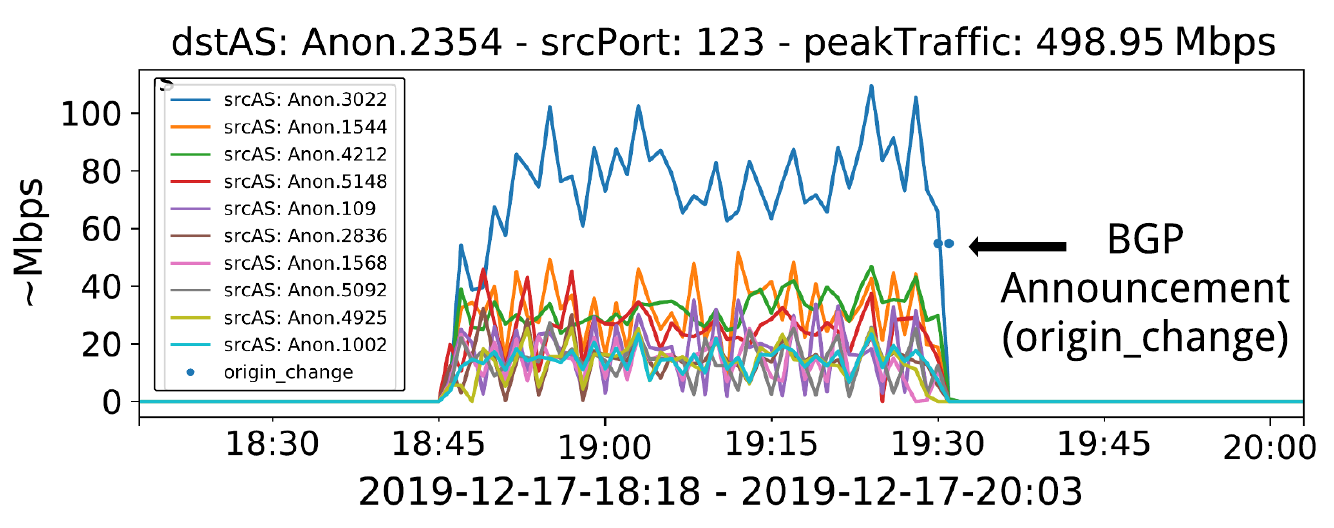}
    \caption{Example of DRDoS attack and traffic re-routing mitigation}
    \label{fig:mitigation}
\end{figure}

\subsubsection{Measuring In-the-Wild Mitigations}

Using the BGP-based analysis explained earlier, we measure whether a mitigation
effort was deployed for the DRDoS attacks detected by \sysname. With respect to
mitigating attacks via traffic re-routing traffic, we found 56 BGP relevant
announcements that occurred during 3 different DRDoS attacks. These BGP
announcements effectively changed the origin AS of IP prefixes owned by the
victim network and redirected traffic to a known traffic scrubbing provider.
All of these mitigation efforts were related to attacks directed towards AS Anon.2354
, with traffic being re-routed to the AS Anon.1890. All 3 attacks for which
mitigation was deployed had a duration greater than 30 minutes. 
With respect to mitigation via BGP-based blackholing, we did not find any
evidence that blackholing was used for remediating any of the DRDoS attacks we
detected.  

Figure~\ref{fig:mitigation} shows an example of DRDoS attack for which a traffic
rerouting mitigation was implemented. As can be seen, in this case the attack
had been ongoing for around 45 minutes, before traffic was re-routed to a
scrubbing center. Traffic re-routing is identified by the BGP announcements to change origin (as seen in Figure~\ref{fig:mitigation}) of the victim prefix to scrubbing center's AS Anon.1890.

While it was a bit surprising that only 3 attacks and only one network operator
used attack mitigation, personal communications with the SoX operators confirmed
that only that one member network made use of a DDoS mitigation plan available
to all of SoX's customers/members. Another surprising observation is that only
long-lived attacks are considered for mitigation. It is possible that one of the
main issue is that currently DDoS detection happens ``manually,'' once the
attack has started to cause noticeable disruption and perhaps network users
start complaining to the operators. Our \sysname system can reduce such
detection delay significantly, by performing DRDoS attack detection in near real
time with an inexpensive open-source solution.

\subsubsection{Improving Attack Mitigation at IXPs}
\label{sec:improving_mitigation}

While re-routing traffic to third-party scrubbing services is a commonly used
strategy for mitigating DDoS attacks, it can become a quite expensive depending
on the size and duration of the attack that a victim is trying to defend
against. Another possibility for mitigation is to rely on IXPs and upstream ISPs
to implement traffic blackholing. However, as explained earlier, currently
blackholing is either an ``all or nothing'' or very coarsely selective strategy
that can cause significant collateral damage~\cite{bgp_blackholing_wild, SelectiveBlackholing}, because it
filters out both legitimate and attack traffic. In this section, we discuss how
\sysname could enable IXPs to help their customer/members who fall victim of
DRDoS attacks, by making traffic blackholing more ``surgical'' so that only
traffic associated with specific services and with specific attack-contributing
source ASes is blocked. This has the potential of significantly reducing
collateral damage.

The strategy we propose is the following. Let $V$ be the victim AS of a DRDoS
attack detected by \sysname, $p$ be the source UDP port abused for reflecting
attack traffic towards $V$, and $S = \{s_1, s_2, \dots, s_n\}$ be the set of
source ASes that send traffic from port $p$ to $V$ during the attack. The IXP
could implement a filtering rule that only blocks all traffic from each source
AS $s_i$ and port $p$ directed towards $V$. Because all information necessary to
create these filtering rules is contained in \sysname's DRDoS alerts, it would
be possible to simply automatically translate each alert into a BGP Flowspec
rule that can be propagated to the IXP's routers, so that the time to implement
the mitigation strategy is greatly reduced, compared to manual intervention.

\begin{figure*}[!ht]
    \begin{subfigure}{0.33\textwidth}
        \centering
        \includegraphics[scale=0.385]{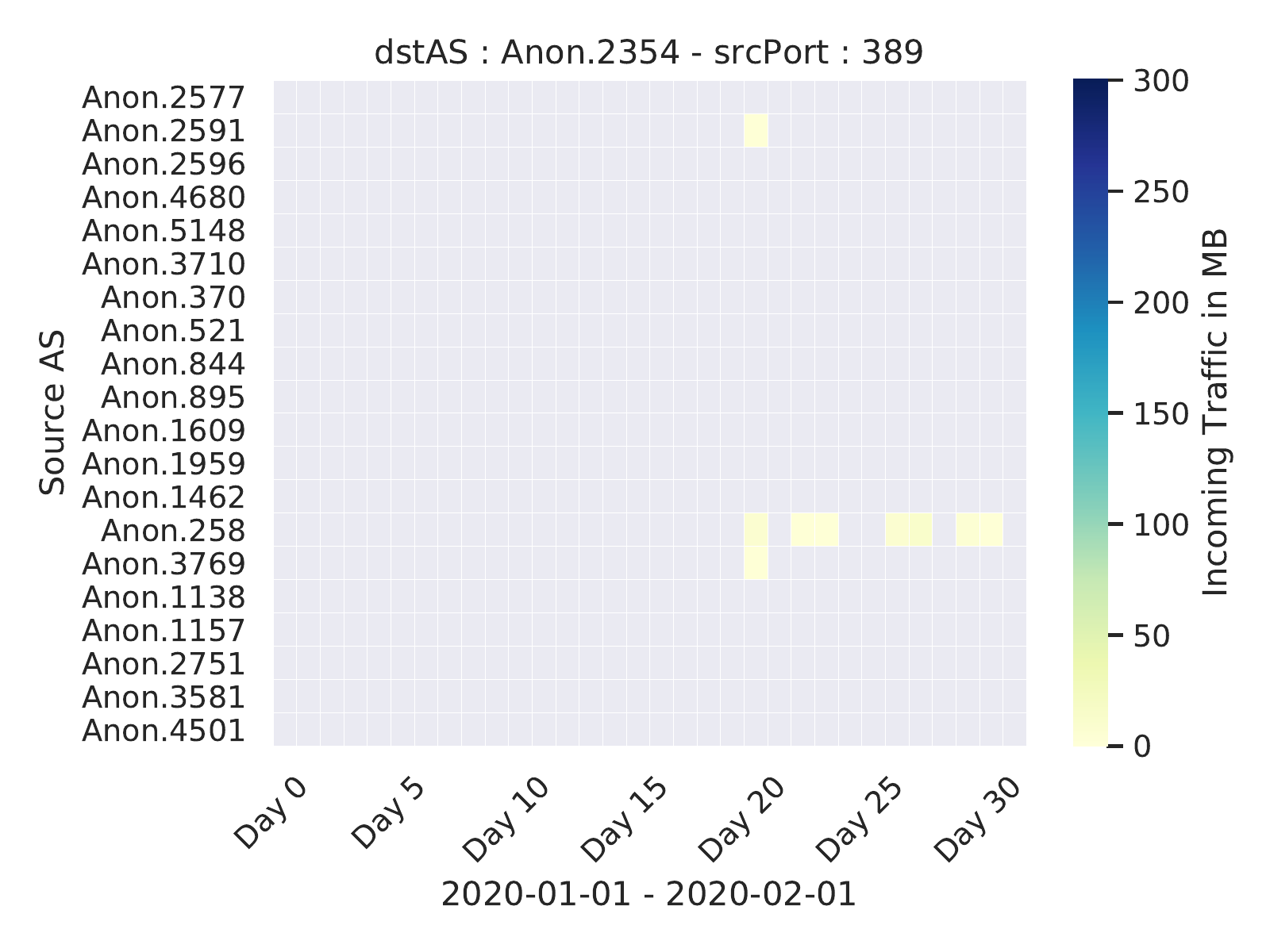}
        \caption{CLDAP}
        \label{fig:heatmap_legit_389}
    \end{subfigure}%
    \begin{subfigure}{0.33\textwidth}
        \centering
        \includegraphics[scale=0.385]{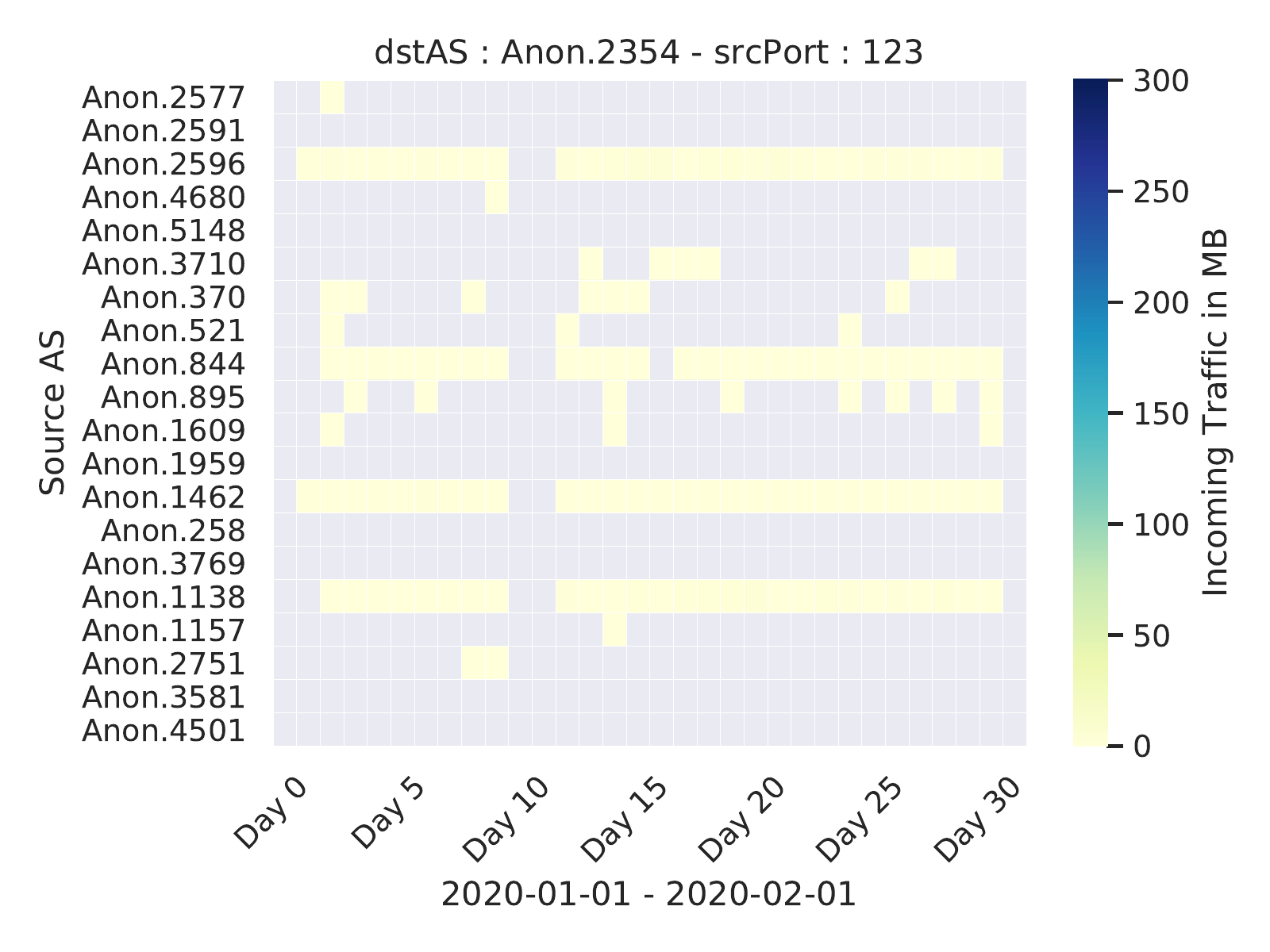}
        \caption{NTP}
        \label{fig:heatmap_legit_123}
    \end{subfigure}%
    \begin{subfigure}{0.33\textwidth}
        \centering
        \includegraphics[scale=0.385]{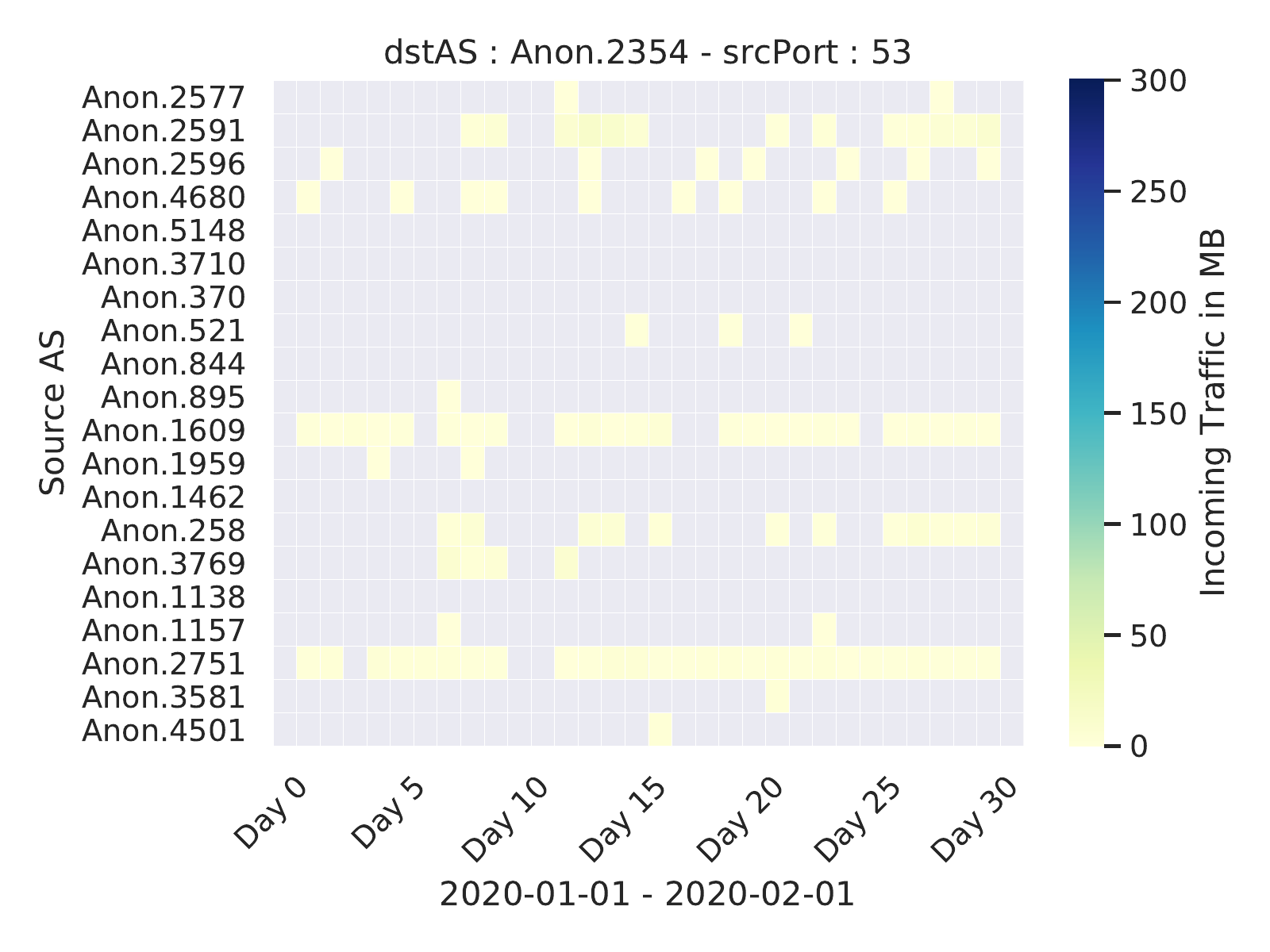}
        \caption{DNS}
        \label{fig:heatmap_legit_53}
    \end{subfigure}
     \begin{subfigure}{0.33\textwidth}
        \centering
        \includegraphics[scale=0.385]{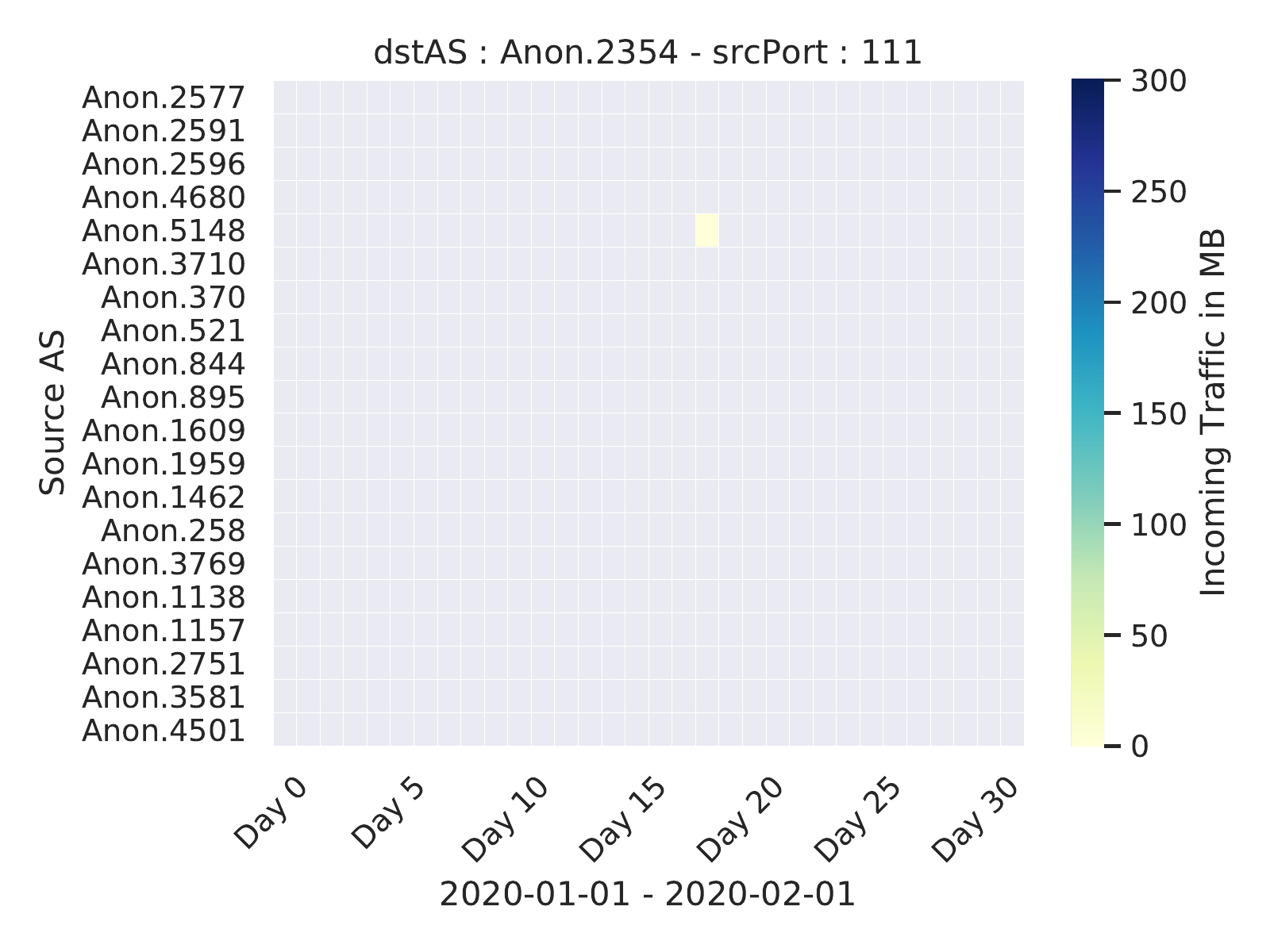}
        \caption{SunRPC}
        \label{fig:heatmap_legit_111}
    \end{subfigure}%
    \begin{subfigure}{0.33\textwidth}
        \centering
        \includegraphics[scale=0.385]{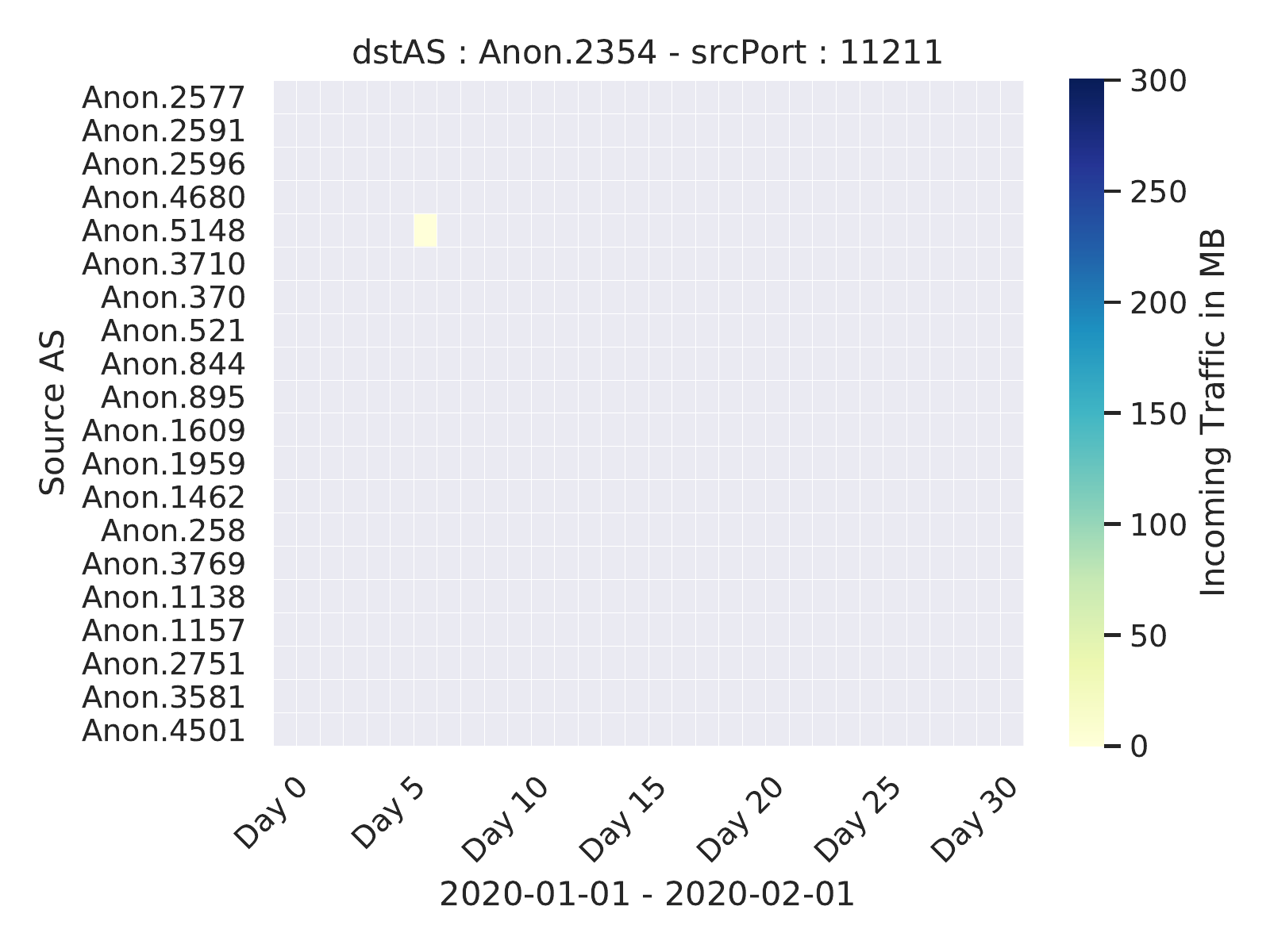}
        \caption{Memcached}
        \label{fig:heatmap_legit_11211}
    \end{subfigure}%
    \begin{subfigure}{0.33\textwidth}
        \centering
        \includegraphics[scale=0.385]{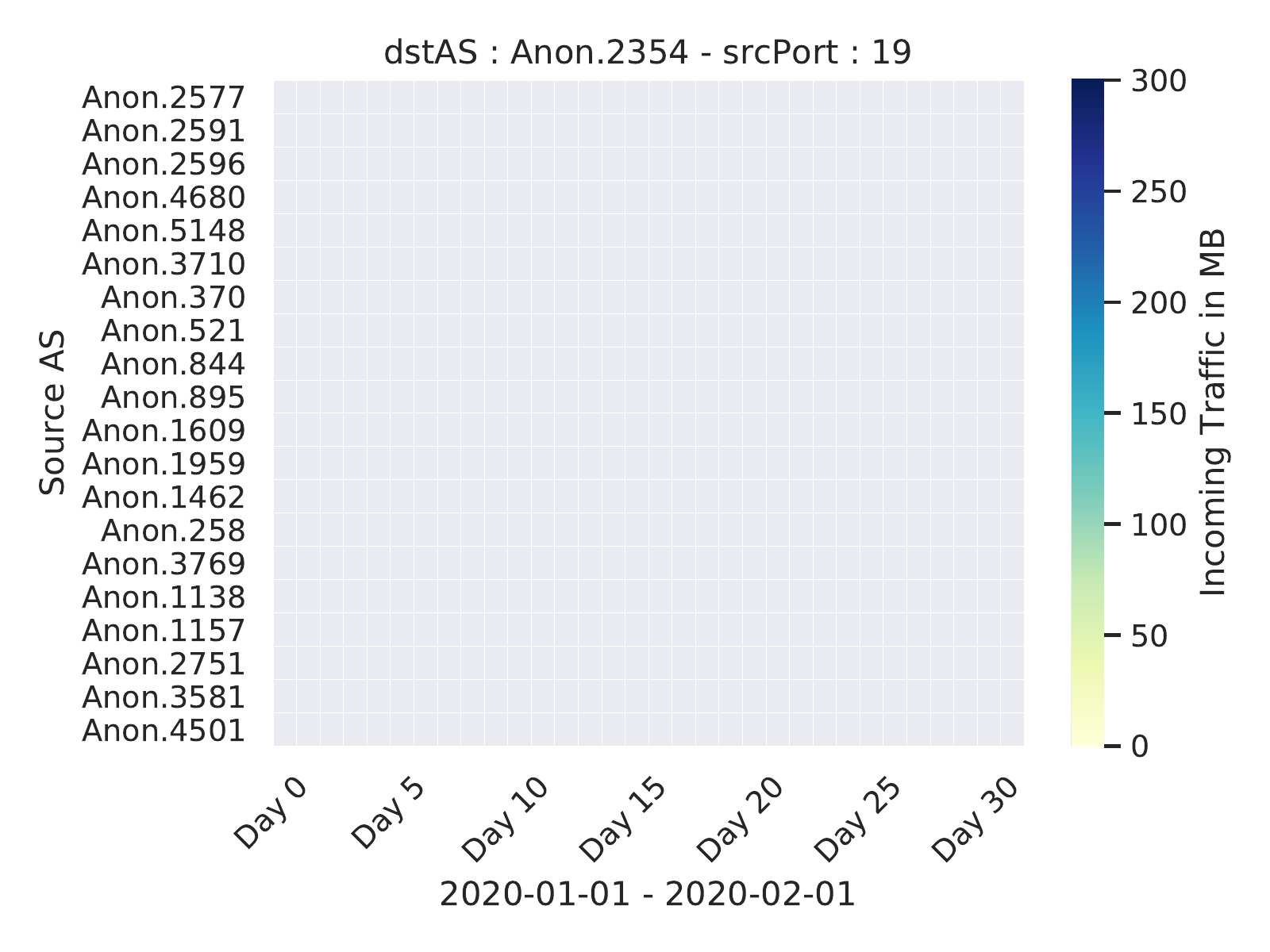}
        \caption{Chargen}
        \label{fig:heatmap_legit_19}
    \end{subfigure}%
    \caption{Daily traffic (in MBytes per day) to destination AS Anon.2354 from a set of 20 legitimate ASes not involved in DRDoS attacks.}
    \label{fig:attack_heatmaps}
\end{figure*}

To understand what is the potential impact to the above mitigation strategy, we
investigate the extent of the ``collateral damage'' (i.e., blocked non-DRDoS
traffic) a target network may incur. To this end, let us consider the traffic
measurements shown in Figure~\ref{fig:attack_heatmaps}. Each heatmap corresponds
to one of the UDP source ports reported in Figure~\ref{fig:ports}, from which we
observed at least one DRDoS attack. All six heatmaps are related to one single
destination AS, which we select as the AS number for which we observed the
largest number of distinct DRDoS attacks, during \sysname's deployment period.
The $x$ axis reports a period of 30 consecutive days of traffic monitoring,
whereas the $y$ axis reports a randomly selected set of 20 source ASes.
These source ASes were selected among all source ASes that during the 30 days
period in the $x$ axis sent at least some traffic from any of the six source UDP
ports. The color of each heatmap cells indicates the total number of MBytes sent
by a source AS to the destination AS during each day. Gray cells indicate zero
bytes, whereas other cell colors indicated the ``intensity'' of the daily
traffic. From all these graphs we exclude attack traffic detected by \sysname.
The reason is that we want to highlight the volume of normal (i.e.,
non-DRDoS attack) traffic typically sent by any source AS to a destination AS,
as seen from the vantage point of an IXP.

Let us consider first Figure~\ref{fig:heatmap_legit_389}, which is related to
CLDAP traffic (port 389). As we can see, it is rare to observe any inter-AS
traffic for this service.
This makes sense, in that CLDAP is primarily meant as an authentication protocol
to be used within a local network. Similarly, ports 19, 111, and 11211 are
unlikely to be used for legitimate inter-AS communication purposes. Therefore,
blocking inter-AS traffic from these ports at the IXP level is unlikely to cause
much collateral damage at all. Services such as NTP (port 123) and DNS (port 53) have a
different traffic profile. Inter-AS traffic in these cases is not uncommon,
though the overall volume can be quite low, and
therefore traffic filtering can produce some observable collateral damage. For example,
filtering all source port 53 traffic towards a destination AS may impact DNS
resolutions for domains whose authoritative name servers are located within the
destination AS. However, let us assume \sysname detects an attack related to
one of these ports/services. A BGP Flowspec rule automatically derived from
\sysname's DRDoS alert would suggest that the IXP filter all traffic coming from
the identified attack source port directed to the victim network. In addition
the filtering rule would specify what source ASes are contributing to the
attack, so that the IXP could block only traffic from a specific source port and
a specific subset of source ASes, thus further limiting possible collateral damage.
Furthermore, filtering could be limited to the duration of the attack. As soon
as \sysname detects that the DRDoS attack is over, a new BGP Flowspec rule could
be issued so that the IXP would stop filtering any traffic towards the target
AS. This approach could help IXPs protect their downstream customer/member
networks from bandwidth exhaustion DRDoS attacks with minimal collateral damage.

\section{Related Work}

In this section we are presenting prior work in the area of DDoS detection and mitigation both in the context of IXPs and large Tier-1 ISPs.  

{\bf Detection:} Sekar et al.~\cite{sekar2006lads} proposes LADS, a multi-stage
flow collection and monitoring infrastructure for DDoS detection at Tier-1 ISPs
that relies on SNMP and Netflow feeds from routers. Unlike LADS, we propose
an open-source detection system designed to be deployed at an IXP to detect and
measuring DRDoS attacks. The specific detection approach using by \sysname is
also different from LADS, as it relies on a combination of online time series
anomaly detection and DRDoS-specific features to minimize possible false alerts. 

Rossow et al.
\cite{AmplificationHell} report on the abuse of amplification protocols in DRDoS
attacks in a large ISP. This work mainly focuses on the abuse of legitimate amplifiers in the
wild and proposes mitigation techniques that should be deployed at the
amplifier level. On the other hand, we focus on building an open-source DRDoS
detection system that can help IXPs in the mitigation of DRDoS attacks against
their customer/member networks.

Hsieh and Chan~\cite{ddos_neural_net} propose a neural networks approach to detect DDoS
attacks. They rely on network features such as number of
packets, number of bytes, time interval variance, packet rate and bit rate.
Similarly, \cite{ddos_ai} proposes to detect DDoS attacks based on Naive Bayes
and Random Forest trees. The drawback of these approaches is that they are not designed for real-time traffic analysis and
deployment at large IXPs. Furthermore, they require large volumes of historical
labeled data for reliable model training, which is often difficult to collect.

{\bf BGP-based mitigation:} Past research \cite{dietzel2018stellar,
giotsas2017inferring} has developed BGP-based techniques that an infrastructure
operator can use to mitigate DDoS attacks. These techniques work in the premise
that a network operator has already deployed a tool to detect DDoS attacks. Once
a DDoS attack is detected then the network operator can inform an upstream
provider, for example, a higher-tier ISP or an IXP, to enforce BGP-based rules
and redirect the attack traffic away from the victim network. The techniques are
primarily based on: a) BGP Blackholing, and b) BGP Flowspec rules.
\cite{bgp_blackholing_wild,Dietzel2016} offer a detailed description and
measurements of the BGP blackholing technique that has become popular and
is offered as a service at many IXPs. \cite{bgp_flowspec} offers an example
application of BGP Flowspec rules. Another study
\cite{NYX,rerouting_defenses,BGP_curtain} proposes an additional BGP-based
technique, called BGP poisoning, to filter out attack traffic. Our work
differs from these approaches because we focus on designing  a detection system
that can be deployed at IXPs to enable the measurement of DRDoS attack
characteristics and that could also be used to enable faster and more selective
attack mitigation. 

{\bf SDN-based mitigation:} Previous approaches have proposed systems that
leverage the capabilities of Software Defined Networking (SDN) technologies and
Network Functions Virtualization (NFV) to detect and mitigate DDoS attacks. To
overcome BGP-based mitigation techniques \cite{bgp_issues,streibelt2018bgp},
Fayaz et al. \cite{ddos_bohatei} propose an OpenDayLight
\cite{medved2014opendaylight} controller and a network of Virtual Machines (VMs)
for increased scalability. The controller is designed to route the traffic
through the VMs to scrub the traffic. Gupta et al. \cite{ixp_sdn, ixp_sdx,
gupta2016industrial} and Dietzel et al. \cite{dietzel2017sdn} have proposed SDN
enabled applications as a network management solution for IXPs. Our approach is
not based on the SDN and NVF paradigms. Instead, our system can complement these
approaches because it can be deployed on infrastructures that do not have
SDN-based capabilities, and could be adapted to work with SDN-based traffic
routing infrastructure at IXPs to mitigate DRDoS attacks in a very selective way
with low collateral damage.  

\section{Conclusion}
In this paper, we studied in-the-wild DRDoS attacks as seen from a large
Internet exchange point (IXP). To enable this study, we first developed
\sysname, an open-source DRDoS detection system specifically designed for
deployment at large IXP-like network connectivity providers and peering
hubs. We then deployed \sysname at Southern Crossroads (SoX), an IXP-like hub that
provides both peering and upstream Internet connectivity services to more
than 20 research and education (R\&E) networks in the South-East United
States. In a period of about 21 months, \sysname detected more than 900
DRDoS attacks towards 31 different victim ASes. An analysis of the
real-world DRDoS attacks detected by our system shows that most DRDoS attacks
are short lived, lasting only a few minutes, but that large-volume,
long-lasting, and highly-distributed attacks against R\&E networks are not
uncommon. We then used the results of our analysis to discuss possible attack
mitigation approaches that can be deployed at the IXP level, before the
attack traffic overwhelms the victim's network bandwidth.
\section*{Acknowledgments}
We would like to thank Brian Flanagan and Scott Friedrich (at the Georgia
Institute of Technology and SoX) for providing the indispensable operational
support and extremely helpful advice on IXmon's deployment. This material is
based in part upon work supported by the National Science Foundation (NSF) under
grant no. CNS-1741608. Any opinions, findings, and conclusions or
recommendations expressed in this material are those of the authors and do not
necessarily reflect the views of the NSF.

\bibliographystyle{unsrt}
\bibliography{bibliography}

\begin{thebibliography}{10}

\bibitem{Kang2013}
Min~Suk Kang, Soo~Bum Lee, and Virgil~D. Gligor.
\newblock The crossfire attack.
\newblock In {\em Proceedings of the 2013 IEEE Symposium on Security and
  Privacy}, SP '13, pages 127--141, Washington, DC, USA, 2013. IEEE Computer
  Society.

\bibitem{EstoniaCyberWar}
Jason Richards.
\newblock Denial-of-service: The estonian cyberwar and its implications for
  u.s. national security.
\newblock \url{http://www.iar-gwu.org/node/65}.

\bibitem{Mirkovic2004}
Jelena Mirkovic and Peter Reiher.
\newblock A taxonomy of ddos attack and ddos defense mechanisms.
\newblock {\em SIGCOMM Comput. Commun. Rev.}, 34(2):39--53, April 2004.

\bibitem{DynDoS}
Kyle York.
\newblock Dyn statement on 10/21/2016 ddos attack.
\newblock \url{http://dyn.com/blog/dyn-statement-on-10212016-ddos-attack/}.

\bibitem{SpamhausDoS}
Matthew Prince.
\newblock The ddos that knocked spamhaus offline.
\newblock
  \url{https://blog.cloudflare.com/the-ddos-that-knocked-spamhaus-offline-and-ho/}.

\bibitem{KrebsDoS}
Brian Krebs.
\newblock Krebsonsecurity hit with record ddos.
\newblock
  \url{https://krebsonsecurity.com/2016/09/krebsonsecurity-hit-with-record-ddos/}.

\bibitem{DigitalAttackMap}
{Digital Attack Map}.
\newblock Top daily ddos attacks worldwide.
\newblock \url{http://www.digitalattackmap.com}.

\bibitem{cloudflare-famousDDoS}
CloudFlare.
\newblock Famous ddos attacks learning objectives.
\newblock \url{https://www.cloudflare.com/learning/ddos/famous-ddos-attacks/}.

\bibitem{aws-DRDoS}
Computer~Business Review.
\newblock {AWS} hit with a record 2.3 tbps ddos attack.
\newblock \url{https://www.cbronline.com/news/record-ddos-attack-aws}.

\bibitem{AmplificationHell}
Christian Rossow.
\newblock {Amplification Hell: Revisiting Network Protocols for DDoS Abuse}.
\newblock In {\em Proceedings of the 2014 Network and Distributed System
  Security (NDSS) Symposium}, February 2014.

\bibitem{Cloudflare-memcached}
CloudFlare.
\newblock Memcrashed - major amplification attacks from udp port 11211, 2018.
\newblock
  \url{https://blog.cloudflare.com/memcrashed-major-amplification-attacks-from-port-11211/}.

\bibitem{NSFocus}
Mina Hao.
\newblock Ddos attack landscape, 2020.
\newblock \url{https://nsfocusglobal.com/ddos-attack-landscape-3/}.

\bibitem{Cloudflare-reflections}
CloudFlare.
\newblock Reflections on reflection (attacks), 2017.
\newblock \url{https://blog.cloudflare.com/reflections-on-reflections/}.

\bibitem{Cloudflare}
{Cloudflare}.
\newblock How cloudflare's architecture allows us to scale to stop the largest
  attacks.
\newblock
  \url{https://blog.cloudflare.com/how-cloudflares-architecture-allows-us-to-scale-to-stop-the-largest-attacks/}.

\bibitem{Akamai}
{Akamai}.
\newblock Why akamai cloud security for ddos protection?
\newblock
  \url{https://www.akamai.com/us/en/solutions/products/cloud-security/ddos-protection-service.jsp}.

\bibitem{NetFlowv9}
CISCO.
\newblock Netflow v9.
\newblock
  \url{https://www.cisco.com/en/US/technologies/tk648/tk362/technologies_white_paper09186a00800a3db9.html}.

\bibitem{bgp_flowspec}
Bgp flowspec.
\newblock
  \url{https://archive.nanog.org/sites/default/files/wed.general.trafficdiversion.serodio.10.pdf}.

\bibitem{SoX}
SoX.
\newblock Southern crossroads.
\newblock \url{https://www.sox.net/}.

\bibitem{description}
European internet exchange association 2012 report on european ixps.
\newblock
  \url{https://www.euro-ix.net/documents/1117-Euro-IX-IXP-Report-2012-pdf}.

\bibitem{DrPeering}
W.B. Norton.
\newblock {\em The Internet Peering Playbook: Connecting to the Core of the
  Internet}.
\newblock DrPeering Press, 2011.

\bibitem{chatzis2013there}
Nikolaos Chatzis, Georgios Smaragdakis, Anja Feldmann, and Walter Willinger.
\newblock There is more to ixps than meets the eye.
\newblock {\em ACM SIGCOMM Computer Communication Review}, 43(5):19--28, 2013.

\bibitem{richter2014peering}
Philipp Richter, Georgios Smaragdakis, Anja Feldmann, Nikolaos Chatzis, Jan
  Boettger, and Walter Willinger.
\newblock Peering at peerings: On the role of ixp route servers.
\newblock In {\em Proceedings of the 2014 Conference on Internet Measurement
  Conference}, pages 31--44, 2014.

\bibitem{ager2012anatomy}
Bernhard Ager, Nikolaos Chatzis, Anja Feldmann, Nadi Sarrar, Steve Uhlig, and
  Walter Willinger.
\newblock Anatomy of a large european ixp.
\newblock In {\em Proceedings of the ACM SIGCOMM 2012 conference on
  Applications, technologies, architectures, and protocols for computer
  communication}, pages 163--174, 2012.

\bibitem{NetFlowL2}
CISCO.
\newblock Netflow layer 2 and security monitoring exports.
\newblock
  \url{http://www.cisco.com/c/en/us/td/docs/ios-xml/ios/netflow/configuration/12-4/nf-12-4-book/nf-lay2-sec-mon-exp.html}.

\bibitem{sFlow}
Peter Phaal and Marc Lavine.
\newblock sflow version 5.
\newblock \url{http://sflow.org/sflow_version_5.txt}.

\bibitem{USCERT-UDPAmp}
{US CERT}.
\newblock {UDP}-based amplification attacks.
\newblock \url{https://www.us-cert.gov/ncas/alerts/TA14-017A}.

\bibitem{routeview}
Routeviews project.
\newblock \url{http://www.routeviews.org/routeviews/ }.

\bibitem{ewmav}
Tony Finch.
\newblock Incremental calculation of weighted mean and variance, 2009.
\newblock \url{https://fanf2.user.srcf.net/hermes/doc/antiforgery/stats.pdf}.

\bibitem{FastNetMon}
Pavel Odintsov.
\newblock Fastnetmon - very fast ddos analyzer with sflow/netflow/mirror
  support.
\newblock \url{https://github.com/pavel-odintsov/fastnetmon}.

\bibitem{CustomerCones}
Matthew Luckie, Bradley Huffaker, Amogh Dhamdhere, Vasileios Giotsas, and
  kc~claffy.
\newblock As relationships, customer cones, and validation.
\newblock In {\em Proceedings of the 2013 Conference on Internet Measurement
  Conference}, IMC ’13, page 243–256, New York, NY, USA, 2013. Association
  for Computing Machinery.

\bibitem{CAIDA_cones}
{CAIDA}.
\newblock As relationships, customer cones, and validation.
\newblock \url{http://data.caida.org/datasets/as-relationships/}.

\bibitem{ip_loc_api}
Ip geolocation mappingk.
\newblock \url{https://ipgeolocation.io/ip-location-api.html}.

\bibitem{drdos_survey}
Fabrice~J. Ryba, Matthew Orlinski, Matthias Wählisch, Christian Rossow, and
  Thomas~C. Schmidt.
\newblock Amplification and drdos attack defense -- a survey and new
  perspectives, 2015.

\bibitem{BlackholingIETF}
T.~King, C.~Dietzel, J.~Snijders, G.~Doering, and G.~Hankins.
\newblock Blackhole bgp community for blackholing.
\newblock \url{https://tools.ietf.org/html/draft-ietf-grow-blackholing-00}.

\bibitem{IANA_reserved}
Network Startup~Resource Center.
\newblock Remote trigger blackhole filtering lab.
\newblock
  \url{https://nsrc.org/workshops/2019/mnnog1/riso/networking/routing-security/en/labs/RTBH-local.html
  }.

\bibitem{bgp_blackholing_wild}
Vasileios Giotsas, Georgios Smaragdakis, Christoph Dietzel, Philipp Richter,
  Anja Feldmann, and Arthur Berger.
\newblock Inferring bgp blackholing activity in the internet.
\newblock In {\em Proceedings of the 2017 Internet Measurement Conference},
  pages 1--14, 2017.

\bibitem{SelectiveBlackholing}
Job Snijders.
\newblock Ddos damage control, cheap and effective.
\newblock
  \url{https://ripe68.ripe.net/presentations/176-RIPE68_JSnijders_DDoS_Damage_Control.pdf}.

\bibitem{sekar2006lads}
Vyas Sekar, Nick~G Duffield, Oliver Spatscheck, Jacobus~E van~der Merwe, and
  Hui Zhang.
\newblock Lads: Large-scale automated ddos detection system.
\newblock In {\em USENIX Annual Technical Conference, General Track}, pages
  171--184, 2006.

\bibitem{ddos_neural_net}
C.~{Hsieh} and T.~{Chan}.
\newblock Detection ddos attacks based on neural-network using apache spark.
\newblock In {\em 2016 International Conference on Applied System Innovation
  (ICASI)}, pages 1--4, 2016.

\bibitem{ddos_ai}
B.~{Zhang}, T.~{Zhang}, and Z.~{Yu}.
\newblock Ddos detection and prevention based on artificial intelligence
  techniques.
\newblock In {\em 2017 3rd IEEE International Conference on Computer and
  Communications (ICCC)}, pages 1276--1280, 2017.

\bibitem{dietzel2018stellar}
Christoph Dietzel, Matthias Wichtlhuber, Georgios Smaragdakis, and Anja
  Feldmann.
\newblock Stellar: network attack mitigation using advanced blackholing.
\newblock In {\em Proceedings of the 14th International Conference on emerging
  Networking EXperiments and Technologies}, pages 152--164, 2018.

\bibitem{giotsas2017inferring}
Vasileios Giotsas, Georgios Smaragdakis, Christoph Dietzel, Philipp Richter,
  Anja Feldmann, and Arthur Berger.
\newblock Inferring bgp blackholing activity in the internet.
\newblock In {\em Proceedings of the 2017 Internet Measurement Conference},
  pages 1--14, 2017.

\bibitem{Dietzel2016}
Christoph Dietzel, Anja Feldmann, and Thomas King.
\newblock Blackholing at ixps: On the effectiveness of ddos mitigation in the
  wild.
\newblock In {\em Passive and Active Measurement: 17th International
  Conference, PAM 2016, Heraklion, Greece, March 31 - April 1, 2016.
  Proceedings}, 2016.

\bibitem{NYX}
J.~M. {Smith} and M.~{Schuchard}.
\newblock Routing around congestion: Defeating ddos attacks and adverse network
  conditions via reactive bgp routing.
\newblock In {\em 2018 IEEE Symposium on Security and Privacy (SP)}, pages
  599--617, 2018.

\bibitem{rerouting_defenses}
M.~{Tran}, M.~S. {Kang}, H.~{Hsiao}, W.~{Chiang}, S.~{Tung}, and Y.~{Wang}.
\newblock On the feasibility of rerouting-based ddos defenses.
\newblock In {\em 2019 IEEE Symposium on Security and Privacy (SP)}, pages
  1169--1184, 2019.

\bibitem{BGP_curtain}
Jared Smith, Kyle Birkeland, Tyler McDaniel, and Max Schuchard.
\newblock Withdrawing the bgp re-routing curtain: Understanding the security
  impact of bgp poisoning through real-world measurements.
\newblock 01 2020.

\bibitem{bgp_issues}
K.~{Butler}, T.~R. {Farley}, P.~{McDaniel}, and J.~{Rexford}.
\newblock A survey of bgp security issues and solutions.
\newblock {\em Proceedings of the IEEE}, 98(1):100--122, 2010.

\bibitem{streibelt2018bgp}
Florian Streibelt, Franziska Lichtblau, Robert Beverly, Anja Feldmann, Cristel
  Pelsser, Georgios Smaragdakis, and Randy Bush.
\newblock Bgp communities: Even more worms in the routing can.
\newblock In {\em Proceedings of the Internet Measurement Conference 2018},
  pages 279--292, 2018.

\bibitem{ddos_bohatei}
Seyed~K. Fayaz, Yoshiaki Tobioka, Vyas Sekar, and Michael Bailey.
\newblock Bohatei: Flexible and elastic ddos defense.
\newblock In {\em Proceedings of the 24th USENIX Conference on Security
  Symposium}, SEC’15, page 817–832, USA, 2015. USENIX Association.

\bibitem{medved2014opendaylight}
Jan Medved, Robert Varga, Anton Tkacik, and Ken Gray.
\newblock Opendaylight: Towards a model-driven sdn controller architecture.
\newblock In {\em Proceeding of IEEE International Symposium on a World of
  Wireless, Mobile and Multimedia Networks 2014}, pages 1--6. IEEE, 2014.

\bibitem{ixp_sdn}
M.~{Chiesa}, C.~{Dietzel}, G.~{Antichi}, M.~{Bruyere}, I.~{Castro}, M.~{Gusat},
  T.~{King}, A.~W. {Moore}, T.~D. {Nguyen}, P.~{Owezarski}, S.~{Uhlig}, and
  M.~{Canini}.
\newblock Inter-domain networking innovation on steroids: empowering ixps with
  sdn capabilities.
\newblock {\em IEEE Communications Magazine}, 54(10):102--108, 2016.

\bibitem{ixp_sdx}
Arpit Gupta, Laurent Vanbever, Muhammad Shahbaz, Sean~P Donovan, Brandon
  Schlinker, Nick Feamster, Jennifer Rexford, Scott Shenker, Russ Clark, and
  Ethan Katz-Bassett.
\newblock Sdx: A software defined internet exchange.
\newblock {\em ACM SIGCOMM Computer Communication Review}, 44(4):551--562,
  2014.

\bibitem{gupta2016industrial}
Arpit Gupta, Robert MacDavid, Rudiger Birkner, Marco Canini, Nick Feamster,
  Jennifer Rexford, and Laurent Vanbever.
\newblock An industrial-scale software defined internet exchange point.
\newblock In {\em 13th $\{$USENIX$\}$ Symposium on Networked Systems Design and
  Implementation ($\{$NSDI$\}$ 16)}, pages 1--14, 2016.

\bibitem{dietzel2017sdn}
Christoph Dietzel, Gianni Antichi, Ignacio Castro, Eder~L Fernandes, Marco
  Chiesa, and Daniel Kopp.
\newblock Sdn-enabled traffic engineering and advanced blackholing at ixps.
\newblock In {\em Proceedings of the Symposium on SDN Research}, pages
  181--182, 2017.

\end{thebibliography}

\appendix
\section{Appendix}

\begin{figure*}[t]
    \centering
    \includegraphics[scale=0.6]{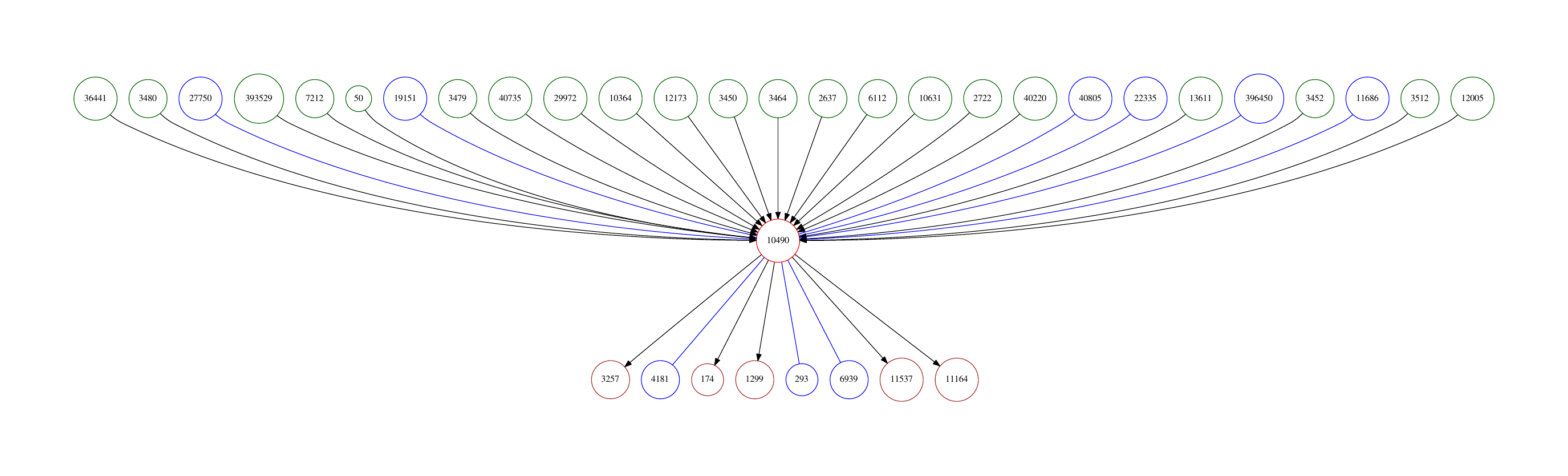}
    \caption{SoX's customers, peers, and providers. Directed (black) edges represent customer-to-provider, whereas undirected (blue) edges represent peering relationships.}
    \label{fig:sox_direct_links}
\end{figure*}

\end{document}